\documentclass[twocolumn,graphics,preprintnumbers,superscriptaddress,amsmath,amssymb]{revtex4-2}
\usepackage{natbib}
\usepackage{dcolumn}
\usepackage{bm}
\usepackage{hyperref}
\usepackage{physics}
\usepackage{mathtools}
\usepackage{float}
\usepackage{amsmath,amsfonts,amsthm}
\usepackage[mathlines]{lineno}
\usepackage{graphicx}
\usepackage{xcolor}
\usepackage{enumitem}

\usepackage{pifont}

\usepackage{verbatim}

\graphicspath{ {./image5/} }

\begin{document}

\title{Phase-space Generalized Brillouin Zone for spatially inhomogeneous non-Hermitian systems}

	\author{Qingya Li}
	\affiliation{Department of Physics, National University of Singapore, Singapore 117551, Singapore}
    
	\author{Hui Jiang}\email{phyhuij@nus.edu.sg}
	\affiliation{Department of Physics, National University of Singapore, Singapore 117551, Singapore}
		
		\author{Ching Hua Lee}\email{phylch@nus.edu.sg}

	\affiliation{Department of Physics, National University of Singapore, Singapore 117551, Singapore}

	\date{\today}

\begin{abstract}
The generalized Brillouin zone (GBZ) has been highly successful in characterizing the topology and band structure of non-Hermitian systems. However, its applicability has been challenged in spatially inhomogeneous settings, where the non-locality of non-Hermitian pumping competes with Wannier-Stark localization and quantum interference, potentially leading to highly non-exponential state accumulation. 
To transcend this major conceptual bottleneck, we develop a general phase-space GBZ formalism that encodes non-Bloch deformations in both position and momentum space, such as to accurately represent spatially inhomogeneous non-Hermitian pumping. 
A key new phenomenon is the bifurcation of the phase-space GBZ branches, which allows certain eigenstates to jump abruptly between different GBZ solutions at various points in real space. The freedom in the locations of such jumps opens up an emergent degree of freedom that protects the stability of real spectra and, more impressively, the robustness of a new class of topological zero modes unique to GBZ bifurcation.
The response from these novel spectral and GBZ singularities can be readily demonstrated in mature metamaterial platforms such as photonic crystals or circuit arrays, where effective real-space hoppings can be engineered in a versatile manner.
Our framework directly generalizes to more complicated unit cells and further hoppings, opening up a vast new arena for exploring unconventional spectral and topological transitions as well as GBZ fragmentation in spatially inhomogeneous non-Hermitian settings.
\end{abstract}
\maketitle

\section{Introduction}

The non-Hermitian skin effect (NHSE) represents a very robust form of localization caused by the directed amplification from asymmetric hoppings. 
Such non-Bloch behavior has led to various paradigm shifts in the way band structures are characterized, as epitomized by modified non-Bloch topological invariants defined on the generalized Brillouin zone (GBZ)~\cite{yao2018edge,yao2018non,yokomizo2019non,lee2019anatomy,zhang2020correspondence,xiong2018does,longhi2019probing,kawabata2020non,lee2020ultrafast,song2019realspace,jiang2023dimensional,wang2024amoeba}. Competition between more than one type of asymmetric hoppings can furthermore modify critical scaling properties~\cite{arouca2020unconventional,liu2024non,li2020critical,qin2023universal,liu2020helical,rafi2022critical,yang2024percolation,yokomizo2021scaling,lin2023topological}and induce non-Hermitian pseudo-gaps~\cite{li2022non} that break the one-to-one correspondence between the open and periodic boundary conditions (OBC and PBC) spectra, challenging existing notions of non-Bloch topological bulk-boundary correspondences~\cite{martinez2018topological,kunst2018biorthogonal,deng2019non,zhu2020photonic,imura2020generalized,yang2020non,guo2021non,yang2022non,shen2024non,yang2024non,ye2025observing,okuma2023non}.

While most existing investigations have focused on NHSE localization against hard open boundaries~\cite{schindler2021dislocation,bhargava2021non,zhang2022universal,lee2019hybrid,kawabata2020higher,okuma2020topological,li2020topological,li2021quantized,li2020topological,sun2021geometric,zhang2023electrical,lee2021many,shen2022non,zhang2022review,shang2022experimental,fang2022geometry,guo2023anomalous,wang2023non,liu2023reentrant,shen2023observation,qin2024dynamical,shen2024enhanced,qin2024kinked,yang2024anatomy,Lei2024,qin2024kinked,xiong2024non}, i.e., physical edges, the underlying directed pumping mechanism gives rise to localized state accumulations as long as translation symmetry is broken. Recently, it has been recognized that spatial inhomogeneities that are not abrupt cut-offs can support other less-known but interesting phenomena such as scale-free eigenstates~\cite{li2021impurity,guo2023accumulation,li2023scale,wang2023scale,xie2024observation,liu2024emergent} and non-Hermitian Anderson localization~\cite{jiang2019interplay,longhi2019topological,longhi2023inhibition,zhai2020many,wang2021anderson,zeng2022real,wang2023observation,acharya2024localization,longhi2024robust}.
These enigmatic phenomena arise because of the nontrivial interplay between the emergent nonlocality from directed non-Hermitian pumping, and the momentum non-conserving spatial inhomogeneities associated with Stark localization, quantum confinement, and band flattening~\cite{liu2020generalized,longhi2020non,longhi2022non,qi2023localization,li2024fate}. 
However, a unified framework encompassing the breadth of these exciting interplays does not yet exist.

To provide this unified framework, we propose in this work the new concept of phase-space GBZs, where the complex momentum deformation for describing the state accumulation profile depends on the spatial position $x$ in addition to the Bloch momentum $k$. By considering the subtle but important skin contributions from spatial hopping gradients, we obtained an analytic ansatz that accurately predicts the energy spectrum and most skin eigenstate profiles in various one-dimensional spatial inhomogeneity profiles, from soft boundaries to impurities to physical edges. This phase-space GBZ construction is also generalized to 2-component systems, where it is shown to accurately predict a new topological phase transition arising from the tuning of the soft boundary width.

Our phase-space GBZ gives a firm theoretical basis for a number of new phenomena. Most salient is the bifurcation of the GBZ into various branches due to spatial inhomogeneity. OBC eigensolutions can exhibit robust discontinuous jumps between them, and as such, acquire non-exponential spatial profiles, distinct from conventional NHSE states. Such ``inhomogeneous skin'' regions also give rise to unconventional but universal spectral branching in the complex plane, beyond what is allowed by spatially homogeneous GBZs. Additionally, in multi-component scenarios, varying the spatial hopping inhomogeneity can also drive topological phase transitions unique to NHSE-pumped states.

\section{Results}

\subsection{1D monoatomic chain with spatially inhomogeneous asymmetric hoppings}

Conventionally, the non-Hermitian skin effect (NHSE) has been rigorously characterized in translationally invariant tight-binding lattices with \emph{constant} but asymmetric hopping amplitudes and hard OBC boundaries~\cite{yao2018edge,kunst2018biorthogonal,yao2018non,lee2019anatomy,song2019non,lee2020unraveling,yokomizo2019non}. Due to this translational invariance, the concept of momentum-space lattice (BZ) remains intact, except that the momenta acquire imaginary contributions, i.e., the GBZ formalism.

In this work, we relax the requirement for translation invariance by modulating the hopping strengths between neighboring sites with a spatially inhomogeneous profile $g(x)$. We first study the minimal (nearest-neighbor) 1D model with a monoatomic unit cell (see Sect. \ref{sec_2b} for more complicated unit cells) under PBCs:
\begin{align}
	H&=\sum_{x=1}^{L-1}g(x)\left(\frac{1}{\gamma}\ket{x}\bra{x+1}+\gamma \ket{x+1}\bra{x}\right)\notag\\
 &+g(L)\left(\frac{1}{\gamma}\ket{L}\bra{1}+ \gamma \ket{1}\bra{L}\right),
 \label{1bH}
\end{align}

on a ring with $L$ sites, where $g(x)/\gamma$ and $g(x)\gamma$ are the left and right hoppings between sites $x$ and $x+1$, respectively.
Here, we have fixed the local hopping asymmetry $\gamma$ to a constant value, since any desired spatial profile of the hopping asymmetry $\gamma(x)$ can be easily obtained from Eq.~\ref{1bH} via a local basis transformation $\ket{x}\rightarrow \gamma^{-x}\Pi_{x'=1}^{x-1}\gamma(x')\ket{x}$, $\bra{x}\rightarrow \bra{x}\gamma^{x}/\Pi_{x'=1}^{x-1}\gamma(x')$, with $\gamma$ set to $\left(\Pi_{x'=1}^{L}\gamma(x')\right)^{1/L}$. 
PBCs are used instead of OBCs, such that the only source of spatial inhomogeneity is from $g(x)$.

\begin{figure}
    \includegraphics[width=\linewidth]{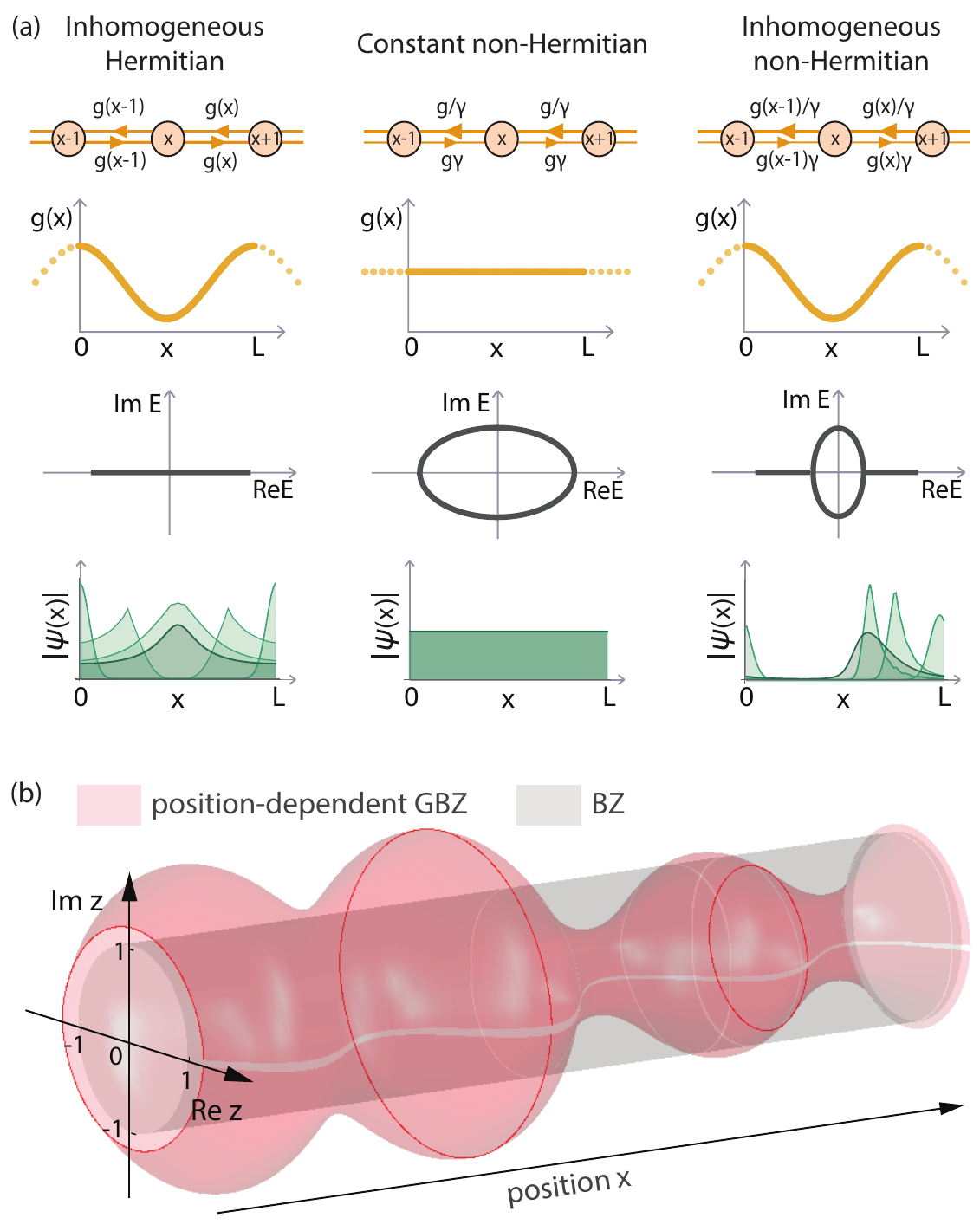}
    \caption{\textbf{Non-trivial Interplay of spatial inhomogeneity and anisotropy in the lattice hoppings. } 
    (a) Left: Spatial inhomogeneity in the lattice hoppings (with profile given by $g(x)$) lead to inhomogeneities in the eigenstates $\psi(x)$ that are locally proportional to $1/\sqrt{g(x)}$. Center: Without spatial hopping inhomogeneity, i.e., PBCs with constant $g(x)$, the eigenstates exhibit constant magnitudes despite non-Hermitian hopping anisotropy $\gamma$ (asymmetric hoppings). Right: With the simultaneous presence of spatial inhomogeneity and asymmetry in the hoppings, new tails branch out from the eigenspectrum and the wavefunctions accumulate asymmetrically against $g(x)$ troughs, which behave like ``partial'' boundaries. 
    (b) An inhomogeneous non-Hermitian hopping strength profile $g(x)$ results in a position-dependent generalized Brillouin zone (GBZ) $ z_{n}(x) = \psi_n(x+1)/\psi_n(x) $ defined in phase space, as given in Eq.~\ref{zdefine}, such that the wavefunctions exhibit different spatial decay or growth rates at different positions. Shown is the GBZ for an illustrative $g(x) = \left(\sin(4\pi x / L) \cdot \cos(2\pi x / L) + 1.2\right)^{-1}$ with $ \gamma = 0.2 $.
    }
    \label{cartoon2}
    \end{figure}

The PBC ansatz Hamiltonian Eq.~\ref{1bH} encompasses the usual well-studied limits as special cases, as we first schematically describe in the following. For $\gamma=1$ [Fig.~\ref{cartoon2}a (Left)], it reduces to a Hermitian nearest-neighbour tight-binding chain with real spectrum and eigenstates $\psi(x)$ that depends only \emph{locally} on the texture $g(x)$. For constant $g(x)=g$ with non-Hermitian $\gamma\neq 1$ [Fig.~\ref{cartoon2}a (Center)], it reduces to the usual Hatano-Nelson model~\cite{gong2018topological,schindler2021dislocation,HN1996prl,HN1997prb,HN1998prb,PhysRevB.92.094204,claes2020skin}  with a complex elliptical spectrum. Even though the eigenstates $\psi(x)$ are pumped leftwards by the hopping asymmetry, they do not have anywhere to accumulate against due to the PBCs and uniform $g(x)$, and thus exhibit spatially uniform amplitudes $|\psi(x)|$.

But with the simultaneous presence of nontrivial hopping asymmetry $\gamma \neq 1$ and non-constant spatial hopping profile $g(x)$ [Fig.~\ref{cartoon2}a (Right)], the non-Hermitian pumped eigenstates are able to accumulate partially against the $g(x)$ inhomogeneities. As schematically shown, illustrative eigenstates generally accumulate towards the right of the $g(x)$ minimum, which is at the center. Notably, due to the directed pumping from right to left, the $g(x)$ minimum behaves like a partial ``boundary'' that non-locally prevents most of the state from occupying the region on its left. However, unlike skin states under genuine OBCs, the spatial state accumulation is manifestly non-exponential, thereby precluding any direct characterization through complex non-Bloch momentum. Also, as compared to the spectrum in Fig.~\ref{cartoon2}a (Center), the inhomogeneity in $g(x)$ has given rise to additional spectral branches at the side, reminiscent of (but distinct from) the OBC spectra of uniform models with asymmetric hoppings~\cite{lee2020unraveling,yang2022designing,tai2023zoology,zhang2024observation,li2020critical,liu2020helical,rafi2022critical,qin2023universal}.

\subsection{Phase-space GBZ for inhomogeneous single-component chains}\label{Result_A}

\subsubsection{General formalism setup}

\noindent To rigorously characterize the anomalous non-local consequences from spatially non-uniform $g(x)$, we examine the Schr\"{o}dinger eigen-equation $H\ket{\psi_{n}}=E_{n}\ket{\psi_{n}}$ of Eq.~\ref{1bH}:
\begin{align}
   \frac{ g(x)}{\gamma }\psi_{n}(x+1)+g(x-1)\gamma \psi_{n}(x-1)=E_{n} \psi_{n}(x),\label{bulk}
\end{align}
where $E_n$ is the eigenenergy of the eigenstate $\ket{\psi_n}=\sum_x \psi_n(x)| x\rangle$. For later convenience, we define the lower and upper bounds of $g(x)$ by $g_\text{min}=\text{Min}[g(x)]$ and $g_\text{max}=\text{Max}[g(x)]$, such that $g_\text{min}=g_\text{max}$ only if the hoppings are completely homogeneous in space.

Since $g(x)$ acts like an energy rescaling factor in the Hermitian continuum limit, local energy conservation requires that $g(x)|\psi_n(x)|^2$ remains invariant.
Extending this to generic non-Hermitian cases where non-local pumping arises from $\gamma\neq 1$, we propose an ansatz for the state amplitude $\psi_{n}(x)$ at site $x$ to be 
\begin{gather}
\psi_{n}(x)=\frac{1}{\sqrt{g(x)}}\gamma^x\prod_{x'=1}^{x}\beta_n(x')\label{ansatz},
\end{gather}
where the $\gamma^x$ keeps track of the pure exponential state accumulation, and $\prod_{x'=1}^{x}\beta_n(x')$ denotes the part of the state accumulation that depends specifically on the $g(x)$ profile, which is the key quantity that we will focus on in this work.

Together, $\gamma$ and $\beta_n(x)$ define the \emph{phase-space} GBZ, as schematically illustrated in Fig.~\ref{cartoon2}b:
\begin{align}
    z_n(x)=\frac{\psi_n(x)}{\psi_n(x-1)}={\sqrt{\frac{g(x-1)}{g(x)}}}\gamma\beta_n(x),\label{zdefine}
\end{align}

which is defined in position-momentum phase space, depending \emph{both} on the position $x$, and the state momentum~\footnote{In more complicated spatially-inhomogeneous lattices with further hoppings, which we will not consider in this work, not only will there be more $\beta_n(x)$ branches, $\gamma$ would also depend on the momentum~\cite{tai2023zoology,lee2020unraveling,lee2022exceptional,zhang2023electrical,lee2019hybrid,lee2019anatomy,li2020topological,kawabata2020higher,okuma2020topological,borgnia2020non,sun2021geometric,li2021quantized,lee2021many,zhang2022review,shen2022non,jiang2023dimensional,qin2023universal}.} indexed through $n$. Different from the usual spatially homogeneous (i.e., constant $g(x)=g$) \emph{OBC} case, where we simply have $z(x)=\gamma$, we have the new $\beta_n(x)$ factor that we call the phase-space GBZ factor, which would compensate for the usual $\gamma$ rescaling factor in a homogeneous PBC system. Its spatial periodicity $\beta_n(x+L)=\beta_n(x)$ is inherited from that of the hopping amplitude profile $g(x)$ and the state amplitude $\psi_n(x)$.

To determine the form of the phase-space GBZ factor $\beta_n(x)$, we substitute this ansatz [Eq.~\ref{ansatz}] into the bulk equation Eq.~\ref{bulk} and arrive at
\begin{gather}
    \frac{\sqrt{g(x)}g(x)}{\sqrt{g(x+1)}}\beta_{n}(x+1)+\frac{{\sqrt{g(x)g(x-1)}}}{\beta_{n}(x)}=E_{n}.\label{Bulk}
\end{gather}
This expression Eq.~\ref{Bulk} applies to generic hopping inhomogeneity $g(x)$, even discontinuous ones to a good approximation (see Sect. \ref{sec_discont_gx}). However, for most cases that we shall consider, we will further make the key assumption of local spatial continuity: If we consider a sufficiently smooth hopping function $g(x)$ such that the spatial gradients satisfy
\begin{align}
    &g'(x)\ll g(x),\label{SmthApprox}\\
    &\beta'_n(x)\ll\beta_n(x),\label{SPT}
\end{align}
in the thermodynamic limit of large $L$, Eq.~\ref{Bulk} simplifies to
\begin{flalign}
    &\beta_{n}(x+1)+\frac{1}{\beta_{n}(x)}=\frac{E_{n}}{ g(x)}&&\label{BulkApprox}\\
     \xRightarrow{\text{continuous GBZ}}\quad &\beta_{n}(x)+\frac{1}{\beta_{n}(x)}=\frac{E_{n}}{g(x)}.&&\label{BulkApproxSmooth}
\end{flalign}
The second line, obtained by assuming local continuity of the phase-space GBZ, decouples the inter-dependency between neighbouring $\beta_n(x)$ and $\beta_n(x+1)$, such that $\beta_n(x)$ can be solved solely from the local hopping $g(x)$ and the eigenenergy $E_n$.
In practice, the locally continuous GBZ assumption can be justified a posteriori by comparing its analytic predictions with numerical diagonalization results. From our results presented later, it turns out that even the rapidly oscillating phases in the wavefunctions do not compromise this assumption, at least for the majority of the reasonably smooth eigenstates.

From Eq.~\ref{BulkApproxSmooth}, the phase-space GBZ factor $\beta_{n}(x)$ can be directed solved in terms of the eigenenergy $E_n$ and the spatial hopping amplitude profile $g(x)$, with a pair of solutions $\beta_{n,\pm}(x)$ given by
\begin{align}
    \beta_{n,\pm}(x)=&\exp\left(\pm \cosh^{-1}\left(\frac{E_{n}}{2g(x)}\right)\right)\notag\\
    =&\frac{E_{n}}{2g(x)}\pm i \sqrt{1-\frac{E_{n}^2}{4g^2(x)}}\ ,\label{PTSol1}
\end{align}
with $|\beta_{n,+}|\geq 1$. To keep track of the extent of spatial state accumulation, we decompose $\beta_{n,\pm}$ as 
\begin{align}
    \beta_{n,\pm}(x)&=\exp\{i[k_{n,\pm}(x)+i\kappa_{n,\pm}(x)]\}\notag\\
    &=e^{ik_{n,\pm}(x)}e^{-\kappa_{n,\pm}(x)},
    \label{skinDepth}
\end{align}
where 
{\small
\begin{align}
    \kappa_{n,\pm}(x)=-\log|\beta_{n,\pm}(x)|=\mp\text{Re}\left[ \cosh^{-1}\left(\frac{E_{n}}{2g(x)}\right)\right]\ 
    \label{skinDepth2}
\end{align}}
represents the local contribution to the \emph{inverse inhomogeneous} skin depth at position $x$, and the phase $k_{n,\pm}(x)$ describes the effective Bloch-like phase oscillations with spatially varying wavenumber $dk_{n,\pm}(x)/dx$.

\subsubsection{Spatially inhomogeneous GBZ branches}

\noindent Although Eq.~\ref{PTSol1} or Eq.~\ref{skinDepth2} may look superficially similar to that of the usual Hatano-Nelson model~\cite{gong2018topological,schindler2021dislocation,HN1996prl,HN1997prb,HN1998prb,PhysRevB.92.094204,claes2020skin}, where $g(x)$ is constant, the inhomogeneity of $g(x)$ brings about various new levels of subtleties. First, labeling the $\beta_{n,\pm}(x)$ solutions such that
\begin{gather}
    |\beta_{n,+}(x)|\geq 1 \geq |\beta_{n,-}(x)|.\label{absBeta}
\end{gather}
We identify the following distinct regions in real space $x$:

\begin{itemize}

\item \textbf{Pure skin region:} $ |\beta_{n,+}(x)|= |\beta_{n,-}(x)|=1$, i.e., $\kappa_{n,\pm}=0$, such that the spatial state profile in these positions is purely exponential (just like usual non-Hermitian skin modes), arising only from the $\gamma^x$ term in Eq.~\ref{ansatz}. It occurs in the region $|\text{Re}(E_n)|\leq 2g(x)$ and $\text{Im}(E_n)=0$. 
\\

\item \textbf{Inhomogeneous skin region:} nonconstant $|\beta_{n,+}(x)|>1 > |\beta_{n,-}(x)|$, such that the spatial state profile is manifestly non-exponential. 
It occurs when $\text{Im}(E_n)\neq0$ or $|\text{Re}(E_n)| >2g(x)$, which represents pockets of weak hopping with no spatially homogeneous analog.

\end{itemize}

Of course, the same physical eigenstate $\psi_n(x)$ can exhibit both pure skin and inhomogeneous skin behaviors at different locations $x$. 
But, whether exhibiting pure or inhomogeneous skin, $\psi_n(x)$ can only incorporate one of the two possible $\beta_{n,\pm}(x)$ solution branches at any particular point $x$. 

We define the choice function $\sigma(x)$ that takes values of $\pm 1$ depending on which branch is chosen at position $x$; exactly how $\sigma(x)$ can be determined will be detailed in the next subsection. Notating the chosen branch as $\beta_n(x)=\beta_{n,\sigma(x)}(x)$ with 
{\small
\begin{align}
    \kappa_{n}(x)=-\log|\beta_{n}(x)|=-\sigma(x)\text{Re}\left[ \cosh^{-1}\left(\frac{E_{n}}{2g(x)}\right)\right],
    \label{skinDepth3}
\end{align}}
we distinguish between two different scenarios for the phase-space GBZ:
\begin{itemize}
\item[\ding{84}] \textbf{Continuous phase-space GBZ:} Either $\beta_n(x)=\beta_{n,+}(x)$ or $\beta_n(x)=\beta_{n,-}(x)$ for all $x$, such that only one branch is ever realized, i.e., $\sigma(x)=\pm 1$ for all $x$. 
\item[\ding{84}] \textbf{Discontinuous phase-space GBZ:} $\beta_n(x)$ switches (jumps) between the $\beta_{n,+}(x)$ and $\beta_{n,-}(x)$ branches at the so-called GBZ inversion points $x_\text{jump}$, which exist due to the spatial inhomogeneity from $g(x)$. 

\end{itemize}

\noindent In a nutshell, the phase-space GBZ connectivity can be classified by the number of $x_\text{jump}$ points where $\sigma(x)$ alternates between $+1$ and $-1$. 
An even number of alternations must occur since $\sigma(x)$ is periodic in $x$ and has to switch an even number of times. A continuous/discontinuous phase-space GBZ corresponds to a zero/nonzero number of $x_\text{jump}$ points -- in this work, we shall explicitly examine only cases with at most two $x_\text{jump}$ points, since more complicated cases can be broken down into multiple discontinuous GBZs in real space and analyzed separately.

\begin{figure*}[!htpb]
    \centering
    \includegraphics[width=\linewidth]{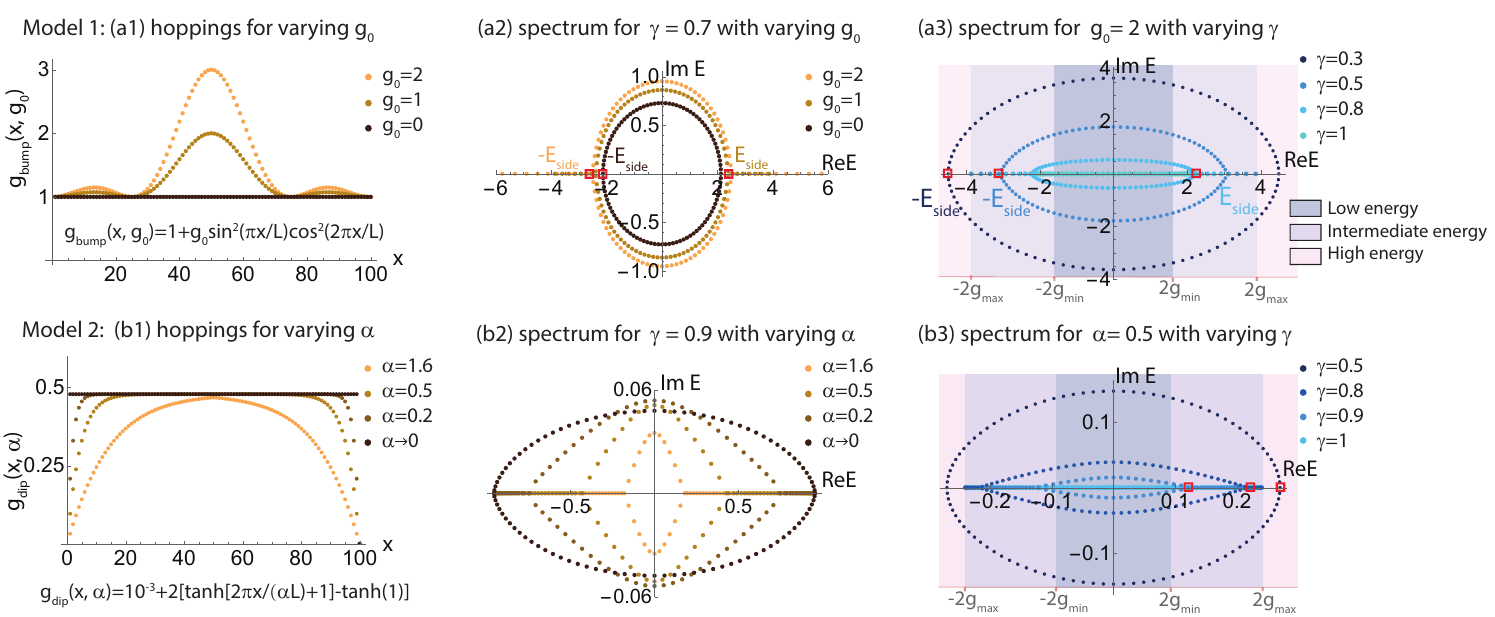}
    \caption{
    \textbf{How the spatial profile $g(x)$ and asymmetry $\gamma$ of the non-Hermitian hoppings affect their energy spectra. } 
    Shown are two illustrative example models with hopping profiles (a1-a3) $g(x)=g_{\text{bump}}(x,g_0)=1 + g_0 \sin^2(\pi x/L)\cos^2(2 \pi x/L)$ and (b1-b3) $g(x)=g_{\text{dip}}(x,\alpha)=10^{-3}+2\left(\tanh(\frac{2 \pi x}{\alpha L} + 1)-\tanh(1)\right)$ for $x\leq L/2$ and symmetric $g(x)$ about $x=\frac{L}{2}$ for $x>L/2$, where $g_0$ and $\alpha$ respectively control the extent of spatial inhomogeneity. (a1,b1) $g(x)$ profiles for these two models. (a2,b2) The corresponding energy spectra at fixed $\gamma$, which branch out into real spectral ``tails'' at branch points $\pm E_\text{side}$ for sufficiently strong inhomogeneity $g_0$ or $\alpha$. (a3,b3) Their corresponding energy spectra at fixed $g_0$ or $\alpha$. As the non-Hermiticity departs from the Hermitian ($\gamma=1$) limit where $E_{\text{side}}=2g_{\text{min}}$, the spectral loops expands and eventually engulfs the real ``tails'' when $E_{\text{side}}$ exceeds $2g_{\text{max}}$. As elaborated in the text, real tails can only exist in the so-called intermediate energy regime where $2g_\text{min}<\text{Re}[E]\leq 2g_\text{max}$.
}
\label{inhomo_figure2}
\end{figure*}

Note that $\kappa_n(x)$ controls the eigenstate amplitude not just locally at $x$, but, in fact, non-locally:
\begin{equation}
|\psi_n(x)|=\frac{\gamma^{x}}{\sqrt{g(x)}} \prod\limits_{x'=1}^{x}e^{-\kappa_{n}(x')},
\label{magpsi}
\end{equation}
which, in terms of the spatial gradient of the state amplitude, takes the form:
\begin{equation}
\frac{\text{d}}{\text{d}x}\left(\log{(}\sqrt{g(x)} \left|\psi_n(x)\right|{)}\right)=\log\gamma -\kappa_n(x).
\label{magpsi2}
\end{equation}

As such, $\kappa_n(x)$ can be interpreted as the spatially dependent correction factor that compensates for the exponential ``pure skin'' accumulation from $\gamma^x$, such that the wavefunction satisfies PBCs. It represents a \emph{position-dependent} reduction of the inverse skin localization length $-\log\gamma$. As the simplest example, consider the spatially homogeneous Hatano-Nelson model where $g(x)=g$ and PBC energy can be
\begin{equation}
E_n=g(\gamma e^{ip_n} + \gamma^{-1}e^{-ip_n})=2g\cosh\left(\log \gamma + i p_n\right),
\label{EnHN}
\end{equation}
where real momentum $p_n\in [0,2\pi)$. We have
\begin{equation}
\kappa_n(x)=\text{Re}\left(\cosh^{-1}(E_n/2g)\right)=\log \gamma,
\end{equation}
which serves to exactly cancel off the $\gamma^x$ accumulation [Eq.~\ref{magpsi}] in the case of the Hatano-Nelson model.

\subsubsection{Determining the allowed energy spectrum}

\noindent The PBC condition $\psi_{n}(x+L)=\psi_{n}(x)$ of the system imposes an important constraint on $\beta_n(x)$ that enables its spectrum $E_n$ and eigenstates $\psi_n(x)$ to be uniquely solved. Substituting the PBC condition into the ansatz Eq.~\ref{ansatz}, we obtain
\begin{align}
   &\gamma^{-L}=\prod_{x=1}^{L}\beta_n(x),\label{GBZBC0}
\end{align}
which, from Eq.~\ref{skinDepth2}, is equivalent to the following handy constraint on the skin depth and eigenenergies:
\begin{align}
    \log(\gamma)=& \frac{1}{L}\sum_{x=1}^{L} \kappa_n(x)\notag\\
    =&-\frac{1}{L}\sum_{x=1}^{L} \sigma(x)\,\text{Re}\left(\cosh^{-1}\left(\frac{ E_n}{2g(x)}\right)\right).
    \label{GBZBC}
\end{align}
In particular, Eq.~\ref{GBZBC} allows the spectrum $E_n$ to be mapped out once $\sigma(x)$ is determined, as will be explained in the following pages. By restricting the 2D complex energy plane to satisfy the constraint equation, the spectrum takes the form of 1D curves or branches, as will be demonstrated later in Figs.~\ref{inhomo_figure2}-\ref{0908fig3}. 

As a corollary to Eq.~\ref{GBZBC}, it is not possible for an entire non-Hermitian system to consist only of pure skin regions (with $\kappa_n(x)=0$), since $\log \gamma\neq 0$. This is just another way of saying that any net exponential NHSE accumulation in a periodic system must be smoothed out and compensated in a $g(x)$-dependent way, as given by Eqs.~\ref{skinDepth3} and \ref{magpsi}.

\subsection{Anatomy of non-Hermitian spectra and eigenstates for spatially inhomogeneous hoppings between monoatomic unit cells}

Fig.~\ref{inhomo_figure2} presents two illustrative models of how the non-Hermitian spectrum behaves as the hopping asymmetry $\gamma$ and spatial inhomogeneity are varied. They are $g(x)=g_\text{bump}(x,g_0)$ and $g(x)=g_\text{dip}(x,\alpha)$, as shown in Figs.~\ref{inhomo_figure2}a1 and~\ref{inhomo_figure2}b1. In the first model, $g_0$ controls the height of the bump at $x=L/2$ [Fig.~\ref{inhomo_figure2}a1]. In the second model, $\alpha$ controls the depth of the dip in the hopping amplitude near the ends $x=1,L$ [Fig.~\ref{inhomo_figure2}b1].  

Even though both hopping profiles result in rather different spectra, they universally behave in qualitatively similar ways [Figs.~\ref{inhomo_figure2}a2,~\ref{inhomo_figure2}b2]. In both cases, the spectrum initially assumes the form of the Hatano-Nelson PBC spectral ellipse [Eq.~\ref{EnHN}] (black) when there is no spatial hopping inhomogeneity, i.e., $g_0=0$ or $\alpha=0$. However, when the spatial inhomogeneity $g_0$ or $\alpha$ is introduced, it generically deforms the spectral loop and saliently introduces real spectral branches or ``tails'' at its sides. We call the eigenenergy branch points where the branches join the loop as $\pm E_\text{side}$ [red square markers in (a2) and (a3)]. These real spectral branches exist only when the non-Hermiticity is not excessively strong, emerging in Figs.~\ref{inhomo_figure2}a3,~\ref{inhomo_figure2}b3 as the hopping asymmetry $\gamma$  decreases towards the Hermitian limit ($\gamma=1$). As will be proven later in this section, these branches can only exist for $2g_\text{min}<|\text{Re}(E)|\leq 2g_\text{max}$ (dubbed the intermediate energy regime), implying that they can only appear when the spectral loop is contained within $|\text{Re}(E)|< 2g_\text{max}$.

Since directed amplification is supposed to continue indefinitely around a homogeneous PBC loop, the appearance of these real eigenenergy branches reveals how spatial inhomogeneity physically behaves like ``partial boundaries'' that can nevertheless completely stop the amplification. For $g_\text{dip}(x,\alpha)$ [Fig.~\ref{inhomo_figure2}b2], it is also interesting that $\text{Max}(\text{Im}(E))$ peaks slightly at moderate values of $\alpha\approx 0.2$, indicative of a slight enhancement of NHSE feedback gain due to hopping inhomogeneity, even though nonzero $\alpha$ corresponds to regions of weak hoppings that should have reduced the overall hopping energies.

\subsubsection{Real spectral branches from GBZ discontinuities}

\noindent To derive the real eigenenergy tail segments that would appear given a generic $g(x)$ and $\gamma\neq 1$, we turn to the boundary condition (Eq.~\ref{GBZBC}), which constrains the set of possible $E_n$ eigensolutions for a given GBZ choice function $\sigma(x)$. Since the full solution set of Eq.~\ref{GBZBC} consists of 1D spectral curves in the complex energy plane, additionally restricting to the real line ($\text{Im}(E)=0$) reduces the solutions to one or more isolated points. In particular, for constant $\sigma(x)=-\text{Sgn}[\log \gamma]$, i.e., continuous phase-space GBZs, the real eigenenergies are found to be $E_n=\pm E_\text{side}$, as defined by
\begin{align}
\qquad\text{Re}\sum_{x=1}^{L} \cosh^{-1}\left(\frac{E_\text{side}}{2g(x)}\right)=L|\log\gamma|.
   \label{sideReE2}
\end{align}

To show that $\pm E_\text{side}$ indeed bounds the two sides of a complex spectral loop [Figs.~\ref{inhomo_figure2}a2,~\ref{inhomo_figure2}a3,~\ref{inhomo_figure2}b2,~\ref{inhomo_figure2}b3], we invoke the following relations between a generic eigenenergy $E_n$ and its $k_n(x)$ and $\kappa_n(x)$ [Eq.~\ref{skinDepth}]:
\begin{align}
    &\frac{\text{Re}(E_{n})}{2g(x)}=\cos(k_{n}(x))\cosh(\kappa_{n}(x)),\label{smoothRe}\\
    &\frac{\text{Im}(E_{n})}{2g(x)}=-\sin(k_{n}(x))\sinh(\kappa_{n}(x)),\label{smoothIm}
\end{align}

which can be obtained by separating the real and imaginary parts of Eq.~\ref{BulkApproxSmooth}. Since $\kappa_n(x)$ cannot identically vanish due to the boundary condition [Eq.~\ref{GBZBC}], $\sin( k_n(x))$ must vanish identically for the real eigenenergy $E_\text{side}$ [Eq.~\ref{smoothIm}]. To continue satisfying Eq.~\ref{GBZBC} for $\text{Re}(E)<E_\text{side}$, $k_n(x)$ in Eq.~\ref{smoothRe} can simply be tuned up; however, doing so inevitably also introduces non-zero $\text{Im}(E)$, as required by Eq.~\ref{smoothIm}. As such, for $\text{Re}(E)<E_\text{side}$, $k_n(x)$ generically generates continuous spectral curves extending into the complex plane.

Most interestingly, if the GBZ choice function $\sigma(x)$ were to vary with $x$, exhibiting jumps between $\pm 1$ values, it is possible to realize a whole continuum of real eigenenergies $E_n$, i.e., the real spectral ``tails'' that all satisfy Eq.~\ref{GBZBC}, as illustrated in Fig.~\ref{0908fig3}a. Previously, with constant $\sigma(x)$, real energies with $E_n>E_\text{side}$ cannot satisfy Eq.~\ref{GBZBC} because the inverse cosh function is monotonically increasing, such that their $\text{Re}\left(\cosh^{-1}\left(E_n/2g(x)\right)\right)$ contributions must exceed that from $E_\text{side}$. However, if $\sigma(x)$ is non-uniform (i.e., exhibits phase-space GBZ discontinuities), jumping at $x=1$ and $x=x_\text{jump}$:
\begin{align}
       \qquad\sigma(x) =\begin{cases}
            1 & 1\leq x< x_\text{jump}\\
            -1 & x_\text{jump}\leq x \leq L ,
        \end{cases}\ ,\label{sigmacases}
\end{align}
the constraint Eq.~\ref{GBZBC} becomes
{\small \begin{align}
\left({\sum_{x=x_\text{jump}}^{L} - \sum_{x=1}^{x_\text{jump}-1}} \right)\text{Re}\left(\cosh^{-1}\left(\frac{E_n}{2g(x)}\right)\right)=L|\log\gamma|,
   \label{GBZBCinv}
\end{align}}
which can be satisfied by a continuum of real $E_n>E_\text{side}$ as the GBZ discontinuity position $x_\text{jump}$ is decreased continuously from $L$.
As shown for an illustrative $g(x)$ [Fig.~\ref{0908fig3}b], such discontinuities are indeed numerically observed [Fig.~\ref{0908fig3}c1] in a typical real-energy state whose absolute energy lies between $|E_{\text{side}}|$ and $|2g_{\text{max}}|$.

In the above, we have established that, due to the freedom in toggling between two different GBZ solutions $\pm \kappa_n(x)$ in a position-dependent manner, as encoded by $\sigma(x)$ jumps, extensively many real energy eigenstates can exist in a PBC system with spatially inhomogeneous hopping amplitudes. This is of profound physical significance because these real energy states do not grow with time, unlike almost every eigenstate in a clean PBC NHSE system~\footnote{For instance, the spectrum of a spatially homogeneous Hatano Nelson model is an ellipse in the complex energy plane, and only two isolated $\pm E_\text{side}=\pm g(\gamma+\gamma^{-1})$ energies are real.}, which has complex energy due to unfettered directional amplification. The juxtaposition of two different GBZs $\kappa_n(x)$ at a spatial GBZ discontinuity $x=x_\text{jump}$ effectively gives rise to an effective spatial ``barrier'' that curtails directional state growth, at least for eigenstates lying in the real spectral branches.

\subsubsection{Spectral behavior in low, high and intermediate energy regimes}

Having discussed how the hopping inhomogeneity leads to the real spectral segments, here we discuss how it affects the full non-Hermitian ($\gamma\neq 1$) spectral behavior in the whole complex plane. To showcase the universality of our arguments, we introduce an additional model with a spatially inhomogeneous hopping profile $g(x)$ that contains a smooth bump at $x=L/2$ and a sharp bump at $x=1$ or $L$ [Fig.~\ref{0908fig3}]. 

Below, we classify the eigenenergies $E_n$ into 3 regimes by comparing $\text{Re}(E_n)$ against the lower and upper bounds of $g(x)$, as colored in Figs.~\ref{inhomo_figure2}a3, \ref{inhomo_figure2}b3 and Fig.~\ref{0908fig3}a: 

\begin{itemize}[leftmargin =.5cm]
\item \textbf{Low energy regime with $|\text{Re}(E_n)|\leq 2g_{\text{min}}$:} 
Only complex $E_n$ allowed, with $\kappa_n(x)\neq 0$ (inhomogeneous skin) across all $x$.

To see why the eigenenergies $E_n$ must be complex, first note that $k_n(x)\neq 0$, because $|\text{Re}(E_n)/2g(x)|\leq 1$ in the LHS of Eq.~\ref{smoothRe}, but $|\cosh(\kappa_{n}(x))|\geq 1$ on the RHS. This is the only possibility because the special case $|\text{Re}(E_n)/2g_{\text{min}}(x)|=1=|\cosh(\kappa_{n}(x))|$ with $|\text{Re}(E_n)|=2g_{\text{min}}(x)$ cannot hold, as $\kappa_n(x)$ cannot identically vanish due to the PBC condition [Eq.~\ref{GBZBC}]. Eq.~\ref{smoothIm} then forces $\text{Im}(E_n)$ to be nonzero.

Note that this low energy regime also exists even when $g(x)$ is spatially uniform, since the above arguments do not involve the details of the $g(x)$ profile, only that $|\text{Re}(E_n)|< 2g_{\text{min}}$. This is already evident from Fig.~\ref{inhomo_figure2}a2, where the low energies form the top of the spectral loops for any $g(x)$ profile. In particular, similar to the spatially homogeneous case, the phase-space GBZ is also continuous with $\kappa_n(x)\neq 0$ everywhere [Fig.~\ref{0908fig3}c3], as required for complex $E_n$ with $k_n(x)\neq 0$ [Eq.~\ref{smoothIm}].

\begin{figure}[!htpb]
    \centering
    \includegraphics[width=\linewidth]{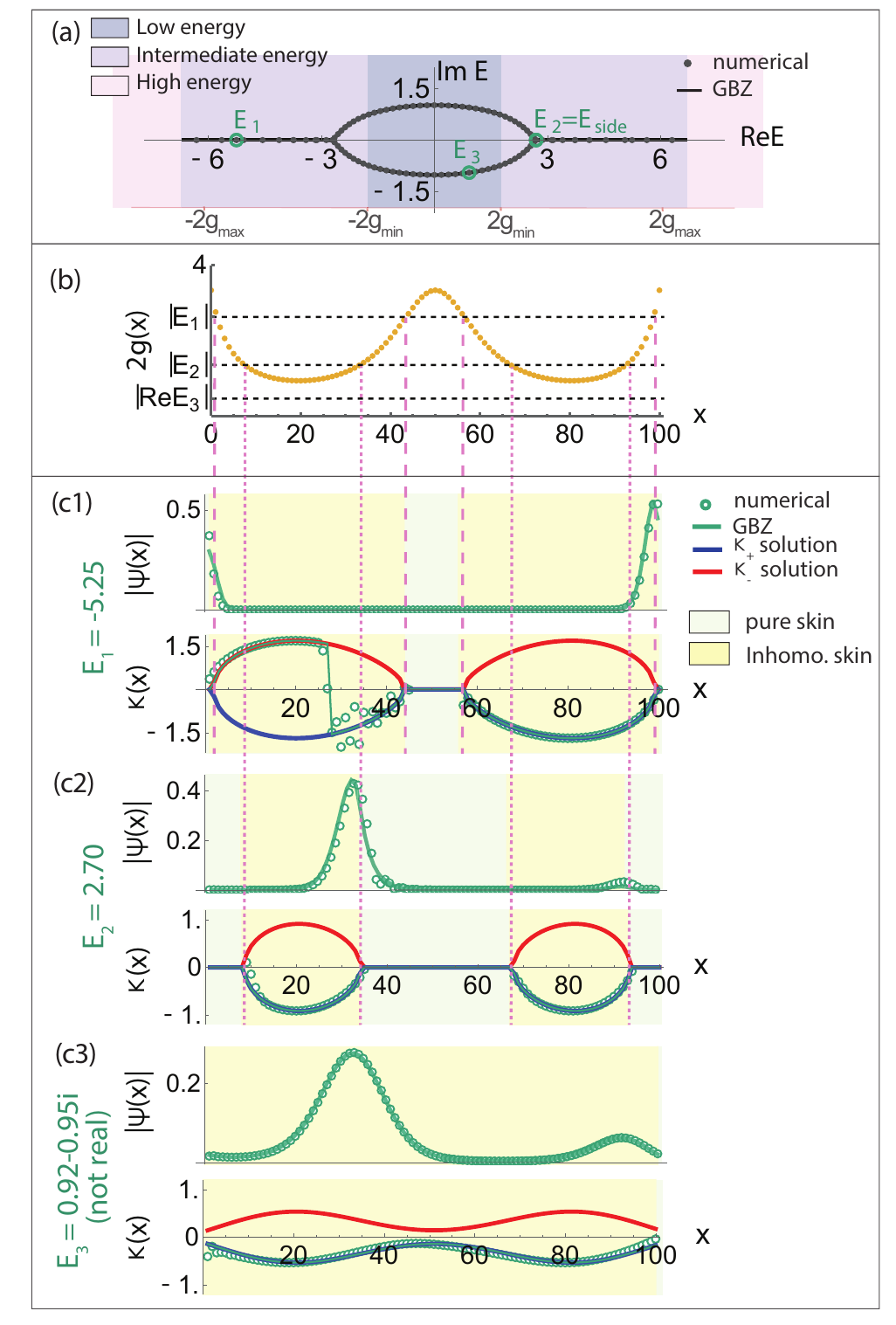}
    \caption{ 
    \textbf{GBZ bifurcations and jumps in the inhomogenous skin regions. }
    (a) Numerically obtained PBC energy spectrum of an illustrative model with $g(x)=1/(\sin(2 \pi x/L) \cos(\pi x/L) + 0.3)$, $L=100$ and $\gamma=0.7$, which satisfies Eq.~\ref{GBZBC}. The branch points $\pm E_\text{side}$ in the intermediate energy regime connect the complex spectral segments with the real ``tails'', whose existence is constrained by Eq.~\ref{sideReE3}. Here, no states exist in the high energy regime. 
    (b) The doubled spatial hopping profile $2g(x)$ is shown against three three illustrative chosen eigenenergy values, which determine the pure and inhomogenous skin regions for their respective eigenstates $\psi$ and GBZs $\kappa(x)$ shown below.
    (c1-c3) The GBZ bifurcates into two branches $\pm \kappa_{n}(x)=\sigma(x)\text{Re}\left(\cosh^{-1}(E_n/2g(x))\right)\neq 0$ at inhomogeneous skin regions (pale yellow) where the hoppings are locally weak, i.e., $\text{Re}(E_n)>2g(x)$ or  $\text{Im}(E_n)\neq0$. The GBZ branch chosen by the numerical eigenstates (green), as computed from Eq.~\ref{magpsi2}, can jump abruptly in the inhomogeneous skin region ($\kappa(x)\neq 0$), as for $E_1$. The jump position $x_{\text{jump}}=27$ is consistent with Eq.~\ref{GBZBCinv}. However, no jump may occur even if GBZ bifurcation occurs, as for $E_2=E_\text{side}$ where the numerically-obtained GBZ adheres to one GBZ solution throughout. No jump can possibly occur when no pure skin region (pale yellow) exists and the GBZ solutions never get to meet, as for $E_3$ not real in the low energy regime.
    }
    \label{0908fig3}
\end{figure}

\item \textbf{High energy regime with $|\text{Re}(E_n)|>2g_{\text{max}}$:}
Either no states at all, or two branches of complex $E_n$ with $\kappa_n(x)\neq 0$ (inhomogeneous skin) meeting at an isolated real eigenenergy point $E_\text{side}$. 

These two possibilities correspond to $\gamma>0.3$ ($\gamma = 0.3$) in Fig.~\ref{inhomo_figure2}a3 and $\gamma>0.5$ ($\gamma = 0.5$) in Fig.~\ref{inhomo_figure2}b3. The argument partially mirrors that of the low-energy regime. Here, $\kappa_n(x)\neq 0$ because $|\text{Re}(E_n)/2g(x)|>1$ in the LHS of Eq.~\ref{smoothRe}, but $|\cos(k_{n}(x))|\leq 1$ on the RHS. With $\kappa_n(x)\neq 0$ for all $x$, Eq.~\ref{smoothIm} forces $\text{Im}(E_n)$ to also not disappear, except for the possible special real solution, which we call $E_n=E_\text{side}$, where $k_n(x)=0$ for all $x$.

\item \textbf{Intermediate energy regime with $2g_{\text{min}}< |\text{Re}(E_n)|\,\leq\, 2g_{\text{max}}$:}

Most interesting is that this intermediate energy regime is characterized by $E_n$ lying between the lower and upper bounds $2(g_\text{min},g_\text{max}]$ of the hopping energy $g(x)$ [Figs.~\ref{inhomo_figure2}a3,~\ref{inhomo_figure2}b3 and Fig.~\ref{0908fig3}a], which has no analogue in the spatially homogeneous limit (where $g_\text{min}=g_\text{max}$). This regime is most intricate because, for a fixed value of $E_n$, there \emph{simultaneously} exist spatial intervals with Re $ E_n>2g(x)$ as well as Re $ E_n\leq 2g(x)$, giving rise to coexisting locally high and low energy regions where $\kappa_n(x)=0$ and $\kappa_n(x)\neq 0$, respectively. 

In particular, it is only in this intermediate energy regime that a continuum of real $E_n$ can exist. 
As shown in Figs.~\ref{0908fig3}a,b, if $2g_\text{min}< E_\text{side}\leq 2g_\text{max}$, the two complex spectral branches from the low-energy regime would meet at $E_n=E_\text{side}$ and continue as a real energy tail that extends till $E_n=2g_\text{max}$, the upper limit of the intermediate energy regime. These real energies in the tail correspond to phase-space GBZ discontinuities at $x=x_\text{jump}$, as given in Eq.~\ref{GBZBCinv}, although nonzero contributions to the sum only come from $\kappa_n(x)\neq 0$ (inhomogeneous skin) regions where $g(x)<E_n/2$.

To understand the role of $E_\text{side}$, note that it is the ``extremal'' solution of the boundary equation [Eq.~\ref{GBZBC}] that still keeps $\sigma(x)$ constant, i.e., $\log(\gamma)= \frac{1}{L}\sum_{x=1}^{L} \kappa_{\text{side}}(x)$. For other real energies $E_n$ with $|E_n|>E_\text{side}$, the constraint $|\kappa_{n}|>|\kappa_{\text{side}}|$ would then require position-dependent $\sigma(x)$ such as to satisfy the boundary equation [Eq.~\ref{GBZBC}]. The upper limit of this continuum of real energies is given by $E_n=2g_\text{max}$, since  $\cosh \kappa_n(x)\leq 1$ and $\sin k_{n}(x)=0$ in Eq.~\ref{smoothRe} and Eq.~\ref{smoothIm}.

As shown in Figs.~\ref{0908fig3}a,b, if $2g_\text{min}< E_\text{side}\leq 2g_\text{max}$, the two complex spectral branches from the low-energy regime would meet at $E_n=E_\text{side}$ and continue as a real energy tail that extends until $E_n=2g_\text{max}$, the upper limit of the intermediate energy regime. These real energies in the tail correspond to phase-space GBZ discontinuities at $x=x_\text{jump}$, as given in Eq.~\ref{GBZBCinv}, although non-zero contributions to the sum only come from $\kappa_n(x)\neq 0$ (inhomogeneous skin) regions where $g(x)<E_n/2$. As shown in Figs.~\ref{0908fig3}c1,c2 for eigensolutions $E_1,E_2$ that lie in the intermediate energy regime [Fig.~\ref{0908fig3}a purple], the jump/s can be numerically extracted from the spatial wavefunction profile via Eq.~\ref{magpsi2}. In Fig.~\ref{0908fig3}c1, the numerical fit to either $\pm\kappa(x)$ solution is good, except at the rather abrupt jump. In Fig.~\ref{0908fig3}c2, there is no jump as the numerical wavefunction adheres to only one GBZ solution branch throughout; in Fig.~\ref{0908fig3}c3, no jump is possible because the two $\kappa(x)$ solutions do not even touch.

However, if $E_\text{side}$ exists in the high-energy regime, i.e., $E_\text{side}>2g_\text{max}$ or
\begin{equation}
\frac{1}{L}\sum_{x=1}^L\cosh^{-1}\left(g_\text{max}/g(x)\right)<|\log\gamma|,
\label{sideReE3}
\end{equation}
there will be no real spectral tail, and the complex spectral branches simply meet at the real point $E_\text{side}$ and terminate there [Figs.~\ref{inhomo_figure2}a3,~\ref{inhomo_figure2}b3]. This would definitely be the case when $g(x)$ is uniform, since $g(x)=g_\text{max}$. Hence Eq.~\ref{sideReE3} gives the threshold for the absence of real eigenenergies: as the hopping asymmetry $\log \gamma$ is increased, directed amplification becomes stronger, and greater spatial inhomogeneity $g(x)$ in the hoppings is needed to stop the amplification and produce asymptotically dynamically stable eigensolutions.

\end{itemize}

\begin{figure*}[!htbp]
    \includegraphics[width=0.7\linewidth]{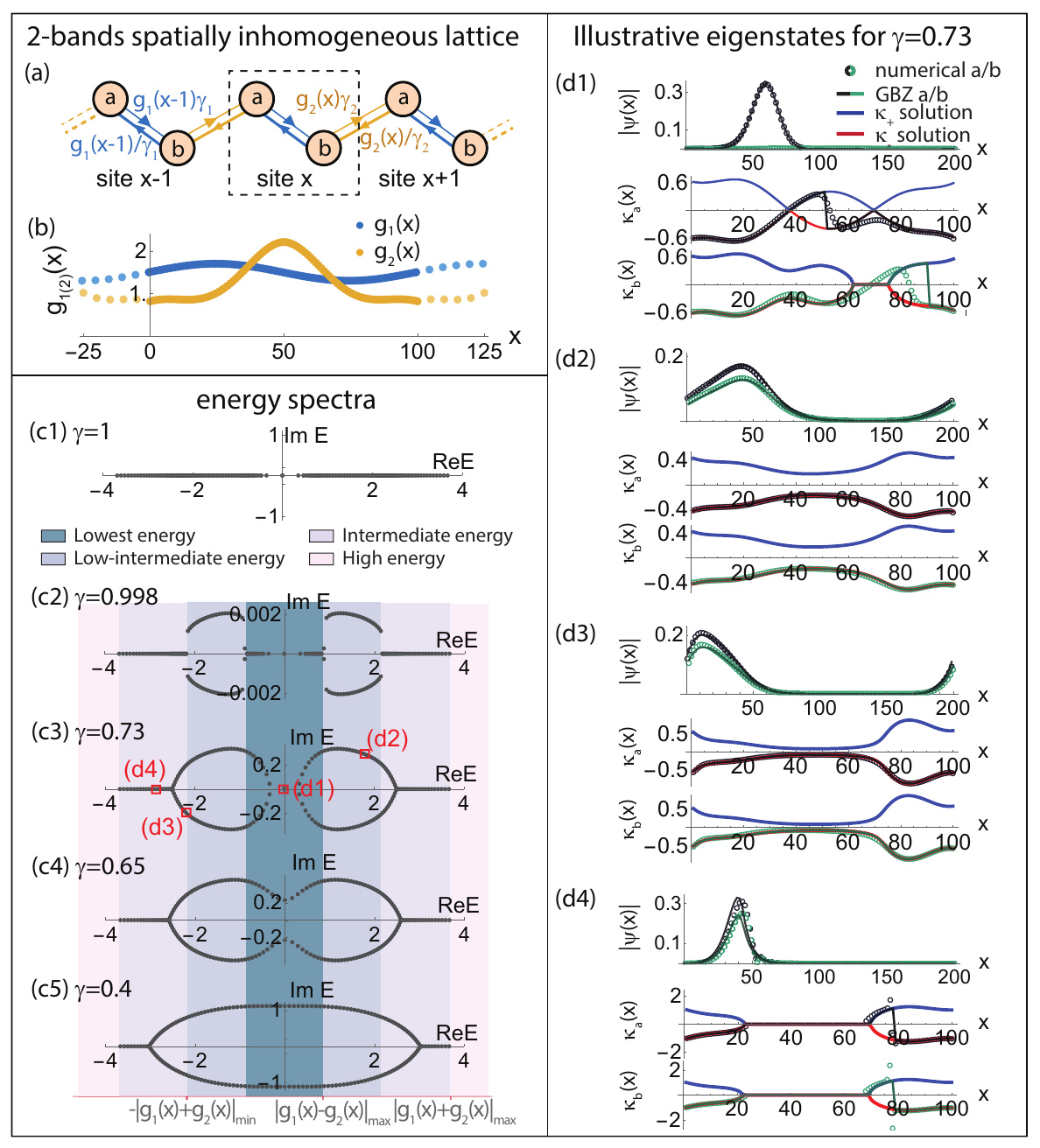}
	\caption{\textbf{The energy regimes and bifurcated GBZs of 2-component spatially inhomogeneous non-Hermitian lattices. }
    (a) Schematic of our spatially inhomogeneous PBC lattice with two atoms per unit cell (dashed), such that the inhomogeneous intra-cell hoppings and inter-cell hoppings have independent profiles $g_1(x)$ and $g_2(x)$.  
    (b) The illustrative profiles used, with $g_1(x)= 1.5 + 0.2 \sin(2 \pi x/L)$ and $g_2(x)=(0.001|(x - L/2)^{1.8}|+ 0.5)^{-1} + 0.2 \cos(4 \pi x/L) + 0.01$, with $L=200$.
    (c1-c5) Energy spectra as the combined hopping asymmetry $\gamma=\gamma_1\gamma_2$ is tuned from the Hermitian $\gamma=1$ limit to $\gamma=0.4$. Different from the 1-component case, there are four distinct energy regimes and two spectral loops that terminate at real ``tails'' in the lowest and intermediate energy regimes. A spectral transition occurs as $\gamma$ is lowered from $\gamma=0.73$ to $0.65$, when the two complex loops coalesce into one, destroying the zero mode. 
    (d1-d4) Wavefunctions of 4 representative numerical eigenstates (black/green for component $a/b$) from (c3), and their approximations by the phase-space GBZs $\pm\kappa_a(x),\pm\kappa_b(x)$. Analogous to 1-component cases, states on the real ``tails'' possess discontinuous phase-space GBZs and exhibit $x_{j,\text{jump}}$ jumps such as to satisfy the PBC constraint Eq.~\ref{2BspecGBZ}. Interestingly, the zero mode of (d1) exhibits completely distinct $\kappa_{a,b}(x)$ profiles both analytically and numerically, a spectacular consequence of the spatial inhomogeneity of the hoppings. 
    }
	\label{Fig4} 
\end{figure*}

\subsection{Phase-space GBZ for inhomogeneous two-component chains}\label{sec_2b}

Here, we generalize previous results on monoatomic unit cells to spatially inhomogeneous lattices with diatomic unit cells, such that the odd and even hoppings have independent spatial profiles $g_1(x)$ and $g_2(x)$. Having a non-trivial unit cell leads to the appearance of multiple spectral bands that not only enrich the phase-space GBZ and complex spectral graphs~\cite{yang2020non,song2019realspace,esaki2011edge,shen2018topological,PhysRevA.97.052115,manna2023inner}, but also host special topological ``soft-interface'' zero modes that have no analogue in spatially homogeneous PBC or OBC systems.
   
A 1D two-component (diatomic unit cell) lattice with spatially inhomogeneous hoppings is defined by

{\small \begin{align}
	H_\text{2-comp}&=\sum_{x=1}^{L}g_1(x)\left(\frac{1}{\gamma_{1}}\ket{x,a}\bra{x,b}+\gamma_{1}\ket{x,b}\bra{x,a}\right)\notag\\&+\sum_{x=1}^{L-1}g_{2}(x)\left(\frac{1}{\gamma_{2}}\ket{x,b}\bra{x+1,a}+\gamma_{2}\ket{x+1,a}\bra{x,b}\right)\notag\\
 &+g_2(L)\left(\frac{1}{\gamma_2}\ket{L,b}\bra{1,a}+\gamma_2\ket{1,b}\bra{L,a}\right),
\end{align}}

as illustrated in Fig.~\ref{Fig4}a, generalizing the well-known (spatially uniform) SSH model. Each diatomic unit cell is indexed by its position $x$, and hoppings across atoms $a,b$ within each unit cell have amplitudes $g_1(x)\gamma_1$ and $g_1(x)/\gamma_1$ in either direction, where $\gamma_1$ is the intra-cell hopping asymmetry. Likewise, hoppings connecting atoms $b,a$ across adjacent unit cells have amplitudes $g_2(x)\gamma_2$ and $g_2(x)/\gamma_2$ in either direction.

The construction of its eigensolutions and phase-space GBZ proceeds analogously to the previous single-component case, albeit with important new features. Writing the $n$-th eigenstate as $\Psi_{n}=\sum_{x}\sum_{ j =a,b}\psi_{n, j }(x)\ket{x, j }$ and substituting it into the time-independent Schr\"{o}dinger eigenequation $H_\text{2-comp}\Psi_{n}=E_{n}\Psi_{n}$ yields the bulk equations 
\begin{align}
    &\frac{g_{1}(x)}{\gamma_{1}}\psi_{n,b}(x)+g_{2}(x-1)\gamma_{2}\psi_{n,b}(x-1)=E_{n} \psi_{n,a}(x),\notag\\
&g_{1}(x)\gamma_{1}\psi_{n,a}(x)+\frac{g_{2}(x)}{\gamma_{2}}\psi_{n,a}(x+1)=E_{n} \psi_{n,b}(x),
\label{2bBulk1}
\end{align}
and is subject to the PBC boundary constraint
 \begin{align}
     \psi_{n, j }(x+L)=\psi_{n, j }(x),\label{2bBound1}
 \end{align}
where $ j =a,b$ labels the sublattice component. Inspired by the success of the single-component ansatz Eq.~\ref{ansatz}, we write down the two-component ansatz for the state amplitude as
\begin{align}
\psi_{n, j }(x)&=\frac{\gamma_1^{x}\gamma_2^x}{\sqrt{g_{1}(x)g_{2}(x)}}\prod_{x'=1}^{x}\beta_{n, j }(x').\label{2bansatz}
\end{align}
Assuming sufficiently smooth odd/even hopping amplitude profiles and local phase-space GBZ continuity:
\begin{align}
    &g'_{1,2}(x)\ll g_{1,2}(x),\label{SmthApprox}\\
    &\beta'_{n, j }(x)\ll\beta_{n, j }(x),\label{SPT}
\end{align}
the bulk equations Eq.~\ref{2bBulk1} for a state with eigenenergy $E_n$ reduce to
\begin{align}
\frac{1}{2}\left(\beta_{n, j }(x)+\frac{1}{\beta_{n, j }(x)}\right)=\omega_{ j }(E_n,x),\label{2bBulk2}
\end{align}
with
\begin{align}
     &\omega_{a}(E_n,x)=\frac{E_n^2-g_{1}^2(x)-g_{2}^2(x-1)}{2\sqrt{g_{1}(x-1)g_{1}(x)g_{2}(x-1)g_{2}(x)}},\notag\\
    &\omega_{b}(E_n,x)=\frac{E_n^2-g_{1}^2(x)-g_{2}^2(x) }{2\sqrt{g_{1}(x)g_{1}(x+1)g_{2}(x-1)g_{2}(x)}}.\label{omega}
\end{align}
Importantly, the equations governing $\beta_{n,a}(x)$ and $\beta_{n,b}(x)$ are completely decoupled, with information of the inter-sublattice couplings entering only through $\omega_{ j =a,b}(E_n,x)$.

In fact, by substituting
\begin{align}\label{replacement}
    &\omega_{ j }(E_n,x)\to\frac{E_{n}}{2g(x)},\qquad \gamma_1\gamma_2\to \gamma,
\end{align}
Eqs.~\ref{omega} for each $ j =a,b$ can be solved in a way identical to the single-component case, even though additional subtleties emerge (as will be discussed later). The intra- and inter-cell hopping asymmetries $\gamma_1,\gamma_2$ appear only through their product $\gamma_1\gamma_2$ because they contribute to non-Hermitian skin pumping successively in a symmetric manner. Thus, the two phase-space GBZ branches can be defined exactly analogously through $\beta_{n, j, \pm}(x)=\exp(ik_{n, j,\pm}(x) -\kappa_{n, j,\pm}(x))$, with $\kappa_{n, j ,\pm}(x)$ and phase $k_{n, j ,\pm}(x)$ being the local contributions to the inverse skin depth and phase:
\begin{align}
&|\beta_{n, j ,\pm}(x)|=\exp( \pm  \cosh^{-1}(\omega_{ j }(E_n,x))),\label{2bSol1}\\
&\kappa_{n, j ,\pm}(x)=\mp \text{Re}\left( \cosh^{-1}(\omega_{ j }(E_n,x))\right).\label{2bSol2}
\end{align} 
As in the 1-component case, the phase-space GBZ may potentially jump discontinuously between the $\kappa_{n, j ,\pm}(x)$ solutions at certain positions $x_\text{jump}$, as encoded in the GBZ choice functions [Eq.~\ref{sigmacases}] $\sigma_{ j }(x)=\pm 1$, depending on whether the eigenwavefunction $ \psi_{n, j }(x)$ assumes the $+1$ or $-1$ branch at $x$. For a given eigenenergy $E_n$, the exact locations of the jumps in $\sigma_{ j }(x)$ can be determined by enforcing the PBC condition
\begin{align}
    \log(\gamma_1\gamma_2)={-}\frac{1}{L}\sum_{x=1}^{L}\sigma_{ j }(x)\text{Re}\left( \cosh^{-1}(\omega_{ j }(E_n,x))\right),
    \label{2BspecGBZ}
\end{align}
such that the eigenstate profile is explicitly
{\small
\begin{align}
&|\psi_{n, j }(x)|\notag\\=&\frac{\gamma_1^{x}\gamma_2^x}{\sqrt{g_{1}(x)g_{2}(x)}}\prod_{x'=1}^{x}\exp\left(\sigma_{ j }(x) \cosh^{-1}(\omega_{ j }(E_n,x))\right).\label{2bandprofile}
\end{align}}

\begin{figure*}[!htpb]
\centering
    \includegraphics[width=\linewidth]{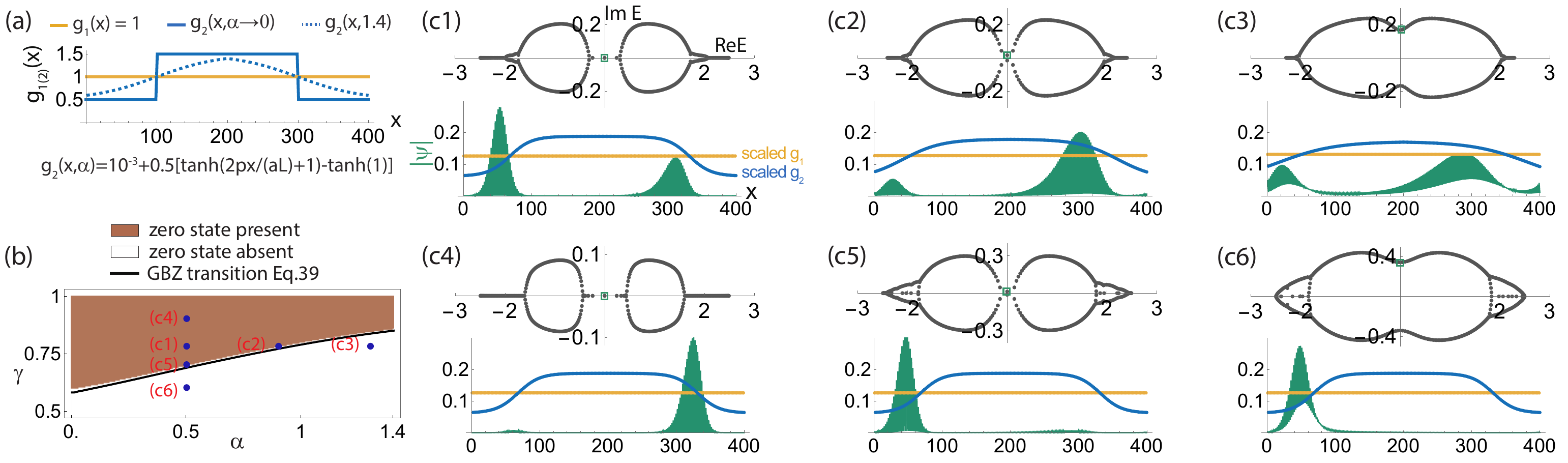}
    \caption{\textbf{Topological phase transition from varying the strength and smoothness of the spatial hopping profile. }
    (a) PBC hopping profile, with uniform intra-unit cell hoppings $g_1(x)=1$ and non-uniform inter-unit cell hopping strengths $g_{2}(x,\alpha)=1+0.5\tanh( 2 \pi \frac{x - \frac{L}{4}}{\alpha L})$, $-\frac{L}{2}<x\leq \frac{L}{2}$, with $L=200$. $g_1(x)$ and $g_2(x)$ intersect to form domain walls at $x=\frac{L}{4}$ and ${3L}{4}$, with wall steepness diverging to form hard boundaries as $\alpha\rightarrow 0$.
    (b) The numerically-determined region (brown) with robust zero modes (numerical tolerance $|E|<0.015$) in the parameter space of $\alpha$ and non-Hermitian asymmetry $\gamma$. It is accurately bounded by the transition curve (black) from Eq.~\ref{PDD2}. 
    (c1-c6) Spectra and corresponding zero mode (or minimal $|E_n|$ state) [Eq.~\ref{2bandprofile}] at illustrative parameter values given in (b). 
    Topological zero modes always occupy only one sublattice (dashed or solid green) per unit cell at the same $x$, just like familiar SSH zero modes, and accumulate against the domain wall intersections $g_{1}(x)=g_2(x)$.
    }
    \label{fig_topo_smoothness}
\end{figure*}

\subsubsection{Real spectra and phase-space GBZ discontinuities for two-component chains}

On first impression, the two-component spatially inhomogeneous problem may seem no more complicated than the 1-component case, since its spectrum can be obtained by replacing $E_n/2g(x)$ with $\omega_{ j }(E_n,x)$, $ j =a,b$ [Fig.~\ref{Fig4}]. However, since $\omega_{ j }(E_n,x)$ has a more complicated spatial dependence, it, in fact, sets more sophisticated and subtle constraints on the spectrum.

Previously, it was established that a real energy continuum requires the presence of phase-space GBZ discontinuities $x_\text{jump}$ [Eq.~\ref{GBZBCinv}], which occur when $|E_n/2g(x)|\leq 1$ for some but not all $x$, i.e., $2g_\text{min}\leq\text{Re}(E_n)\leq 2g_\text{max}$ (the intermediate energy regime). For this 2-component case, a real spectral continuum thus requires that $|\omega_{ j }(E_n,x)|\leq 1$ for some but not all $x$, such that Eq.~\ref{2BspecGBZ} can be satisfied for a continuum of real $E_n$ by continuously adjusting $x_\text{jump}$ in $\sigma_{ j }(x)$. Substituting this condition into Eqs.~\ref{omega} 
and assuming sufficiently smooth spatial inhomogeneities such that $g_{1,2}(x)\approx g_{1,2}(x\pm 1)$,
{\small
\begin{align}
         |g_1(x)+g_2(x)|_{\text{min}}&<|\text{Re}(E_n)|<|g_1(x)+g_2(x)|_{\text{max}}\label{Eside2b2}\\
          \text{or}\qquad\qquad 0& <|\text{Re}(E_n)|\leq |g_1(x)-g_2(x)|_{\text{max}}.
    \label{Eside2b}
\end{align}}
Unlike the 1-component case where $g_1(x)=g_2(x)=g(x)$, real energies can also exist within an additional low(est) energy regime $0 <|\text{Re}(E_n)|< |g_1(x)-g_2(x)|_{\text{max}}$ [Eq.~\ref{Eside2b}], where $|\text{Re}(E_n)|$ is not larger than the hopping amplitude difference between the odd and even bonds. This new regime obviously does not exist in the 1-component case where Eq.~\ref{Eside2b2} simply reduces to the intermediate energy regime. As such, the spectral plane is divided into up to 4 different energy regimes:

\begin{itemize} [leftmargin = 0.5cm]
\item \textbf{Lowest energy regime with $|\text{Re}(E_n)|< |g_1(x)-g_2(x)|_{\text{max}}$: } Real energies are possible.
\item \textbf{Low-intermediate energy regime with $|g_1(x)-g_2(x)|_{\text{max}}\leq |\text{Re}(E_n)|\leq |g_1(x)+g_2(x)|_{\text{min}}$:} Only complex energies are possible; however, this regime may not exist for sufficiently dissimilar $g_1(x),g_2(x)$.
\item \textbf{Intermediate energy regime with $|g_1(x)+g_2(x)|_{\text{min}}<|\text{Re}(E_n)| \leq |g_1(x)+g_2(x)|_{\text{max}}$:} Real energies are possible.
\item \textbf{High energy regime with $|g_1(x)+g_2(x)|_{\text{max}}<|\text{Re}(E_n)|$:} Complex energy branches (if any) join up along the real line.
\end{itemize}
In particular, continua of real energies can exist in two distinct scenarios: in the intermediate energy regime, which exists only for spatially inhomogeneous systems (similar to the 1-component case), or in the lowest energy regime, where the contrast between adjacent $g_1(x),g_2(x)$ hoppings is sufficient to block directed amplification.

Shown in Fig.~\ref{Fig4} is an illustrative two-component model [Fig.~\ref{Fig4}a] with intra- and inter-component hopping inhomogeneity profiles $g_1(x),g_2(x)$ that intersect at two different locations [Fig.~\ref{Fig4}b]. As non-Hermiticity is introduced and $\gamma$ departs from unity [Fig.~\ref{Fig4}c], spectral loops appear across the lowest to intermediate energy regimes (cyan to light purple). Even though real energies are allowed in the lowest energy regime (cyan), the complex loops join up to form one loop with sufficiently strong non-Hermitian asymmetry $\gamma$, a phenomenon with no single-component analog.

It is instructive to examine the GBZ profiles of the representative eigenstates indicated in Fig.~\ref{Fig4}c3 by red hollow squares. As presented in Fig.~\ref{Fig4}d, only the real energy eigenstates (d1) and (d4) possess pure skin regions where at least one GBZ component is degenerate, i.e., $\kappa_a(x)=0$ or $\kappa_b(x)=0$. These degeneracies allow for ``hidden'' switching of the actual GBZ branches adopted by the numerical eigenstates $\psi$, which in turn gives rise to spectacular GBZ jumps, i.e., discontinuities in $\sigma_j(x)=\pm 1$ necessary for satisfying Eq.~\ref{2BspecGBZ}. While both GBZ components $\kappa_{a,b}(x)$ look identical in (d2)-(d4), they can become completely distinct when the spatial gradient $g_2(x)-g_2(x-1)$ dominates over the eigenenergy $E_n$ in $\omega_{a/b}(E=0,x)$ of Eq.~\ref{omega}, as for the zero mode of (d1).

\subsection{Topological transitions from spatial inhomogeneity}

Most interestingly, spatial hopping inhomogeneity in a 2-component lattice can also drive topological phase transitions. Ordinarily, in a non-Hermitian SSH model with uniform $g_1(x)=g_1$ and $g_2(x)=g_2$ hopping amplitudes, it is well-known that a topological phase boundary occurs at ``domain walls'' where they swap. Here, with spatially inhomogeneous $g_1(x)$ and $g_2(x)$, it is reasonable to expect that  $g_1(x)=g_2(x)$ intersections still function as topological interfaces, since they demarcate the regions $g_1(x)<g_2(x)$ and $g_2(x)>g_1(x)$ that are supposed to represent different phases.

\begin{figure*}[!htpb]
\centering
    \includegraphics[width=\linewidth]{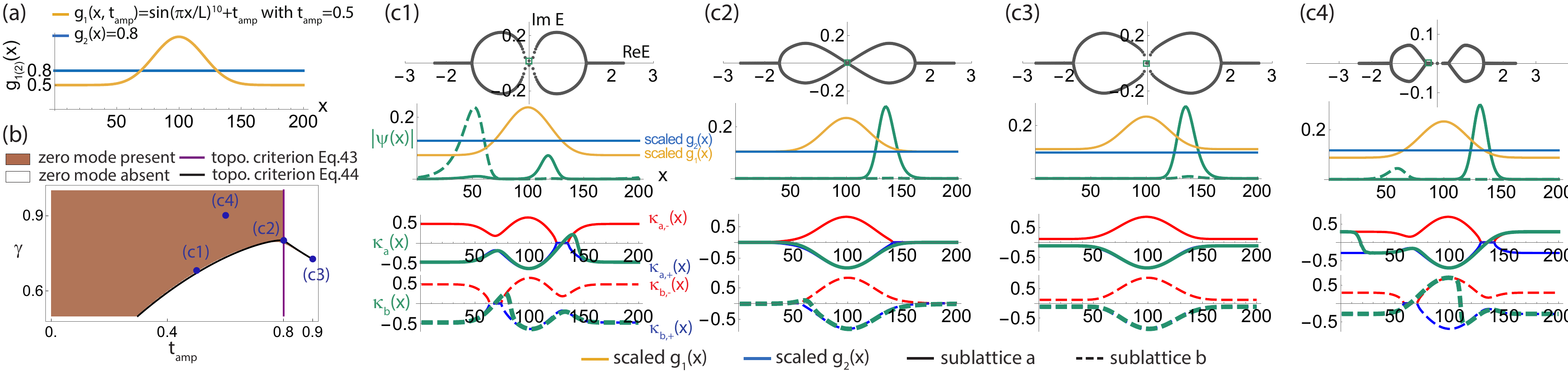}
    \caption{\textbf{Topological phase transition and the crucial role of GBZ jumps. } 
    (a) The PBC hopping profile, as given by constant $g_2(x)$ and inhomogeneous $g_1(x,t_{\text{amp}}) =\sin(\pi x/L)^{10}+t_{\text{amp}}$, $L=200$, with the offset $t_{\text{amp}}$ controlling whether $g_1(x)$ and $g_2(x)$ intersect to form topological domain walls.    
    (b) The numerically determined region hosting topological zero modes (brown, with numerical tolerance $|E|<0.02$) in the parameter space of $\gamma$ and $t_\text{amp}$. It is accurately demarcated by Eq.~\ref{PDD22} (purple) and Eq.~\ref{PDD2} (black), which gives the phase-space GBZ threshold boundaries between continuous and discontinuous GBZs. 
    (c1-c4) Numerically-obtained spectra, state profiles [Eq.~\ref{2bandprofile}] and GBZ occupancies of illustrative $E=0$ zero modes in (b). Deeper into the topological regime (c1,c4), the zero modes are always localized near the ``soft boundary'' defined by $g_1(x) = g_2(x)$ and occupy only one sublattice (dashed or solid green) per unit cell, even though the other sublattice may be occupied away from the soft boundary. However, unlike the bulk zero mode (c3), only the topological states in (c1) and (c4) have discontinuous $\kappa_{j}(x)$ jumping between two branches of $\kappa_{j,\pm}(x)$. 
    }
    \label{fig_topo_amplitude}
    \end{figure*}

However, PBC spatial inhomogeneity complicates the stability of zero modes in various ways. Firstly, among two non-empty regions separated by a domain wall, one of them must already possess non-trivial bulk topology in a bipartite system. Secondly, the bulk is now spatially inhomogeneous, such that its GBZ description rightly lives in phase space and not just momentum space. Thirdly, since phase space encompasses position coordinates, the shape of the domain wall itself affects the topology. Indeed, as will be shown in Figs.~\ref{fig_topo_smoothness} and \ref{fig_topo_amplitude}, the topological modes are not fully confined to $g_1(x)=g_2(x)$ interfaces but, in fact, penetrate nonlocally and nonexponentially into other parts of the system.

\subsubsection{Topological criteria}

Here, we describe a new type of topological robustness protected by GBZ bifurcations. Unlike topological modes at open boundaries, which are simply protected by bulk topological invariants, the inhomogeneous PBC isolated zero modes (i.e., in Fig.~\ref{Fig4}d1) turn out to be crucially protected by GBZ jumps. These jumps facilitate the realization of real spectra (which includes $E_n=0$), as discussed in the paragraphs surrounding Eqs.~\ref{GBZBCinv} and \ref{2BspecGBZ}, and are also further elaborated in Sect.~\ref{sec:single_state}.

As such, to have robust isolated zero modes (and not just trivial $E=0$ crossings), the following two criteria on $\gamma$ and $g_1(x),g_2(x)$ must be satisfied:
\begin{align}
& |g_1(x)-g_2(x)|_{\text{min}}=0,\label{PDD22}\\
 \& \quad   &|\log(\gamma)|< |\log(\Tilde{\gamma}_{\text{topo}})|,\label{PDD2}  
\end{align}
where 
{\small
\begin{align}
    &\Tilde{\gamma}_{\text{topo}}=\exp\left( \frac{1}{L}\sum_{1<|\omega_{ j }(E=0,x)|}\cosh^{-1}|\omega_{ j }(E=0,x)|\right)\label{sideReE2zero},
\end{align}}

is the $\gamma$ threshold that separates continuous and discontinuous GBZ scenarios,with $j=a$ or $b$ yielding negligible differences in the phase diagram (see Sect.~\ref{sec:single_state}). 

The criterion $|g_1(x)-g_2(x)|_{\text{min}}=0$ [Eq.~\ref{PDD22}] is simply that the intra- and inter-unit cell hopping profiles  $g_1(x)$ and $g_2(x)$ intersect to form spatial ``domain walls'' that, by construction, have one side with nontrivial bulk topology. Its implications for the bifurcated GBZ are as follows: when Eq.~\ref{PDD22} is satisfied and $g_1(x)$ and $g_2(x)$ are sufficiently smooth, $\kappa_{n=0,j,\pm}(x)\approx 0$ [Eqs.~\ref{omega},~\ref{2bSol2}] for some $x$ and $j=a,b$. Since $\kappa_{n=0,j,\pm}\neq 0$ for most other $x$, Eq.~\ref{PDD22} can hence be understood as the requirement for the simultaneous existence of both inhomogeneous and pure skin regions in the zero eigenmode. 

It is also insightful to recast the criterion $|\log(\gamma)|< |\log(\Tilde{\gamma}_{\text{topo}})|$ [Eq.~\ref{PDD2}] into a much more intuitive approximate form when $g_1(x),g_2(x)$ are sufficiently smooth. Neglecting spatial gradients in Eq.~\ref{omega}, 
\begin{align}
\kappa_{n, j ,\pm}(x)&={\mp}\text{Re}\left(\cosh^{-1}|\omega_{ j }(E=0,x)|\right)\notag\\
&\approx{\mp} \text{Sgn}[g_1(x)-g_2(x)]\log\frac{g_1(x)}{g_2(x)},\label{PDD4}
\end{align}
which is just the inverse decay length of SSH-type topological modes without additional skin localization, valid when the inhomogeneities in $g_1(x),g_2(x)$ contribute only ``adiabatically''. Eq.~\ref{PDD4} simplifies $\Tilde{\gamma}_{\text{topo}}$ (for $\Tilde{\gamma}_{\text{topo}}<1$) to
\begin{eqnarray}
\Tilde{\gamma}_{\text{topo}}&=&e^{-\frac{1}{L}\sum_{x=1}^L\kappa_{n, j ,-}(x)}\notag\\
&\approx &\left(\prod_{g_2(x)>g_1(x)}^L\frac{g_2(x)}{g_1(x)}\prod_{g_2(x)<g_1(x)}^L\frac{g_1(x)}{g_2(x)}\right)^{\frac{1}{L}}\notag\\
&=&e^{\left\langle\left|\log\frac{g_2(x)}{g_1(x)}\right|\right\rangle},
\label{PDD3}
\end{eqnarray}
where the average $\left\langle\left|\log\frac{g_2(x)}{g_1(x)}\right|\right\rangle$ is taken over all $x\in[1,L]$. As such, for sufficiently smooth $g_1(x),g_2(x)$, the criterion Eq.~\ref{PDD2} reduces to
\begin{equation}
|\log\gamma|<\left\langle\left|\log\frac{g_2(x)}{g_1(x)}\right|\right\rangle,
\label{PDD5}
\end{equation}
which places an upper bound on the non-Hermitian hopping asymmetry $|\log\gamma|$, above which it pumps all states across the inhomogeneous PBC chain and destroys the zero modes. This result applies exclusively for ``soft spatial boundaries'' with no OBC analog: it gives the upper bound for the eigenfunction to be ``patched up'' to satisfy the PBC condition [Eq.~\ref{2BspecGBZ}] through appropriately placed GBZ jumps in $\sigma_j(x)$. Note that the criterion Eq.~\ref{PDD2} (or Eq.~\ref{PDD5}) is trivially satisfied in the Hermitian limit of $\log\gamma=0$, in which only the criterion Eq.~\ref{PDD22} exists.

\subsubsection{Illustrative examples for the topological criteria}

In Figs.~\ref{fig_topo_smoothness} and Fig.~\ref{fig_topo_amplitude}, we showcase numerical topological phase diagrams of two illustrative models and how their phase boundaries are accurately determined by the criteria given in Eqs.~\ref{PDD22} and \ref{PDD2}. We also present some illustrative eigenstates [Eq.~\ref{2bandprofile}] and highlight the characteristic amplitude profiles and GBZ jumps of isolated zero eigenmodes.

Fig.~\ref{fig_topo_smoothness} presents a spatial hopping profile [Fig.~\ref{fig_topo_smoothness}a] whose topological phase boundary [Fig.~\ref{fig_topo_smoothness}b] is well approximated by Eq.~\ref{PDD2} alone. It features a uniform $g_1(x)=1$ (yellow) and a $g_2(x)$ (blue) which cuts $g_1(x)$ at domain walls $x=\frac{L}{4}$ and $x=\frac{3L}{4}$. The wall steepness is controlled by a parameter $\alpha$; as $\alpha\rightarrow 0$, the steepness diverges, leading to an OBC-like hard boundary. The phase boundary curve matches excellently with results from numerical diagonalization (brown) and indicates that, as the domain wall becomes softer (with larger $\alpha$), the zero mode becomes more fragile, being more easily destroyed as $\gamma$ moves away from the Hermitian limit $\gamma=1$. 

Even though all eigenstates (green) are localized to some extent in a spatially inhomogeneous setting, topological zero modes characteristically occupy only one sublattice, as plotted in Figs.~\ref{fig_topo_smoothness}c1,c4. As the topological line gap closes and forms a point gap [Figs.~\ref{fig_topo_smoothness}c2,c3,c5,c6], the eigenstates start to disperse away from the domain walls, and both sublattices assume nonzero occupancy.

In Fig.~\ref{fig_topo_amplitude}, we present a different spatial hopping profile [Fig.~\ref{fig_topo_amplitude}a] such that its topological phase boundary [Fig.~\ref{fig_topo_amplitude}b] is demarcated by both Eqs.~\ref{PDD22} and \ref{PDD2}. As the parameter $t_\text{amp}$ increases, the region where $g_1(x)$ (yellow) is larger than $g_2(x)$ (blue) broadens until it finally occupies the whole system and no domain wall exists. This scenario is exactly demarcated by the purple line, to the right of which $|g_1(x)-g_2(x)|_\text{min}=0$ [Eq.~\ref{PDD22}] no longer holds. 

That the zero modes are fundamentally protected by GBZ jumps can be seen in Fig.~\ref{fig_topo_amplitude}c, which showcases three illustrative eigenstates on the threshold boundary given by $|\log \gamma|=|\log\Tilde\gamma_\text{topo}|$ [Eq.~\ref{PDD2}]: (c1) within the topological phase, (c2) at the boundary given by Eq.~\ref{PDD22}, (c3) outside the topological phase, as well as a reference (c4) deep within the topological phase.
Evidently, numerical GBZ jumps (green) in their GBZ branches of blue and red curves confirm that the topological nature of the $E=0$ state corresponds to the presence of a discontinuous jump between $\kappa_{j,\pm}(x)$ and the simultaneous presence of both inhomogeneous and pure skin regions.

\subsection{Discussion}

In this work, we have formulated a new theoretical framework for generically treating the interplay of the NHSE and spatial lattice hopping inhomogeneity, as encoded by $\gamma$ and $g(x)$, respectively. This is a subtle scenario because the spatially non-uniform energy scale not only competes with NHSE accumulation through Wannier-Stark localization but also distorts the skin accumulation and deforms the effective lattice momentum in a position-dependent manner. 

Central to our formalism is the phase-space generalized Brillouin zone (GBZ), which captures the effective non-Bloch deformation in both position and momentum space. For any PBC eigensolution $E_n$, the phase-space GBZ bifurcates into two possible solution branches within regions of relatively weak hoppings $2g(x)<E_n$, leading to non-exponential ``inhomogeneous skin'' state profiles. Crucially, discontinuous jumps in the adopted GBZ branch give rise to an emergent degree of freedom that results in real ``tails'' in the energy spectrum. Physically, these real eigensolutions represent states that are prevented from indefinite growth by spatial inhomogeneity.

Two-component settings encompass new forms of topological robustness as different $g(x)$ components intersect to form spatial domain walls. The real spectral solutions from GBZ jumps can also exist at very low energies, including the topological zero modes in particular. Unlike the well-known topological edge modes, these isolated zero modes are protected by the phase-space GBZ bifurcations, lending their robustness from the emergent freedom in the GBZ jump positions. As shown both theoretically and numerically, such topological phase boundaries can be accurately predicted through our criteria given by Eqs.~\ref{PDD22} and \ref{PDD2}.

By generalizing Eqs.~\ref{2bansatz} and \ref{omega}, our phase-space GBZ framework can be extended to inhomogeneous systems with arbitrarily many components, such that the GBZ of each component depends non-linearly on the inter-component hoppings. Additionally, generalizing the spatially inhomogeneous hoppings beyond nearest neighbors replaces Eq.~\ref{2bBulk2} with a higher-degree Laurent polynomial that splits the GBZ solutions into multiple branches. In all, these are expected to generate far more intricate GBZ jumps, leading to many more hidden degrees of freedom that stabilize new emergent spectral branches, some possibly containing topological modes with higher symmetry~\cite{Chen2013, Chiu2016, Wen2017, Chen2011}. Further generalization to higher dimensions and multiple interacting boundary conditions could lead to significant new subtleties in the already fragmented GBZ structure, opening up a vast playground for future research into non-Hermitian localization.

Experimentally, spatially inhomogeneous non-Hermitian systems are as accessible as their usual uniform lattice counterparts. Non-Hermitian lattice models have already been realized in electrical circuits~\cite{lee2018topolectrical,helbig2019band,Wang2020,hofmann2020reciprocal,zou2021observation,su2023simulation,hohmann2023observation,zhang2023electrical,zhang2024observation,zou2024experimental,guo2024scale},
 cold atoms~\cite{li2019, ren2022,liang2022dynamic,zhou2022}, 
 photonics~\cite{song2020two,  yu2021nonhermitian, song2023nonhermitian, parto2020non,lin2021optics, zhong2021nonhermitian}, 
 programmable quantum simulators~\cite{kamakari2022,koh2023measurement,peng2020simulating,chertkov2023characterizing,Shen2023,Shen2024,koh2024realization,koh2022prl,okuma2022nonnormal}, 
 and mechanical/acoustic systems~\cite{yang2022nonhermitian, braghini2021nonhermitian, jin2022exceptional}. In most metamaterial platforms, the effective hopping strengths can be spatially tuned in a versatile manner -- for instance, the individual components of an electrical circuit array can be tuned at will, with effectively asymmetric couplings simulated using operational amplifiers~\cite{lee2018topolectrical,Ezawa2019c,Ezawa2019d}. The phase space GBZ can be reconstructed from the eigenstate profiles, which can for instance be obtained in electrical circuits through impedance measurements~\cite{Imhof2018,Lu2023,Zhang2022nonHermitian,Franca2024} alongside the resonance spectrum.

\section{Methods}

\begin{figure*}[!htbp]
    \includegraphics[width=\linewidth]{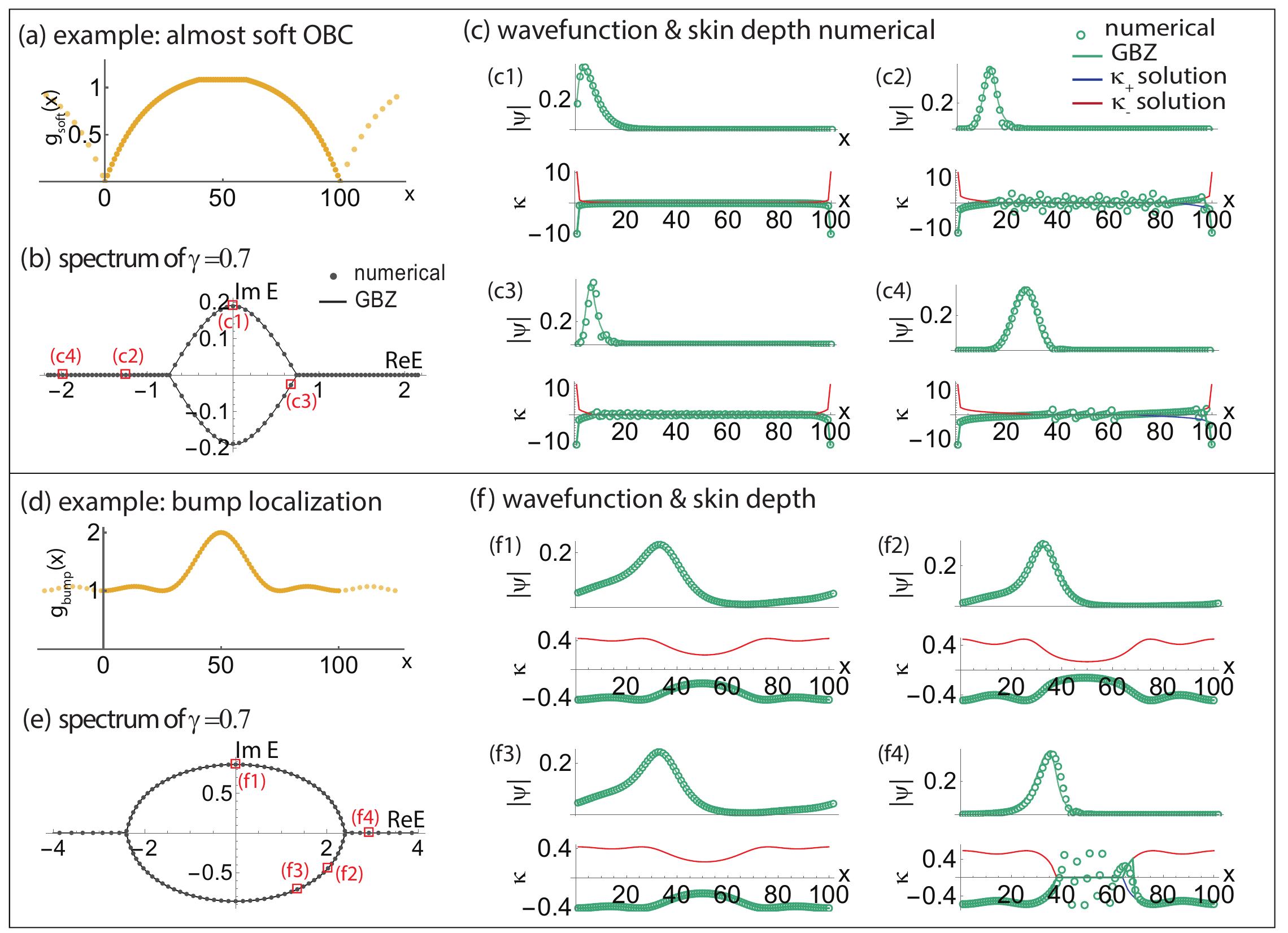}
	\caption{\textbf{Phase-space GBZ: further examples. } Our phase-space GBZ approach well-approximates both energy spectra [Eq.~\ref{sideReE2}] and wavefunctions [Eq.~\ref{ansatz} and Eq.~\ref{PTSol1}] in smooth 1D single-band inhomogeneous lattices. 
    Two different inhomogeneous hopping profiles are chosen, $g(x)=g_{\text{soft}}(x)$ (a-c) from Eq.~\ref{osci_exp} and $g(x) =g_{\text{bump}}(x)$ (d-f)
 from Eq.~\ref{osci_exp}, with $\gamma=0.7$. 
 The numerical results agree almost perfectly with the GBZ solutions, not just in the spectra (b,e), but also the wavefunctions (green) as well as their fits with both continuous and discontinuous GBZ regions (blue and red in c,f). Some fluctuations in the numerically reconstructed GBZ are unavoidable in the pure skin region of finite inhomogeneous systems, as in the discussions in Sec.~\ref{pure_fluc}, but they effectively average to zero in the phase-space GBZ.
        }
	\label{fig7A} 
\end{figure*}

\begin{figure*}[!htbp]
    \includegraphics[width=0.8\linewidth]{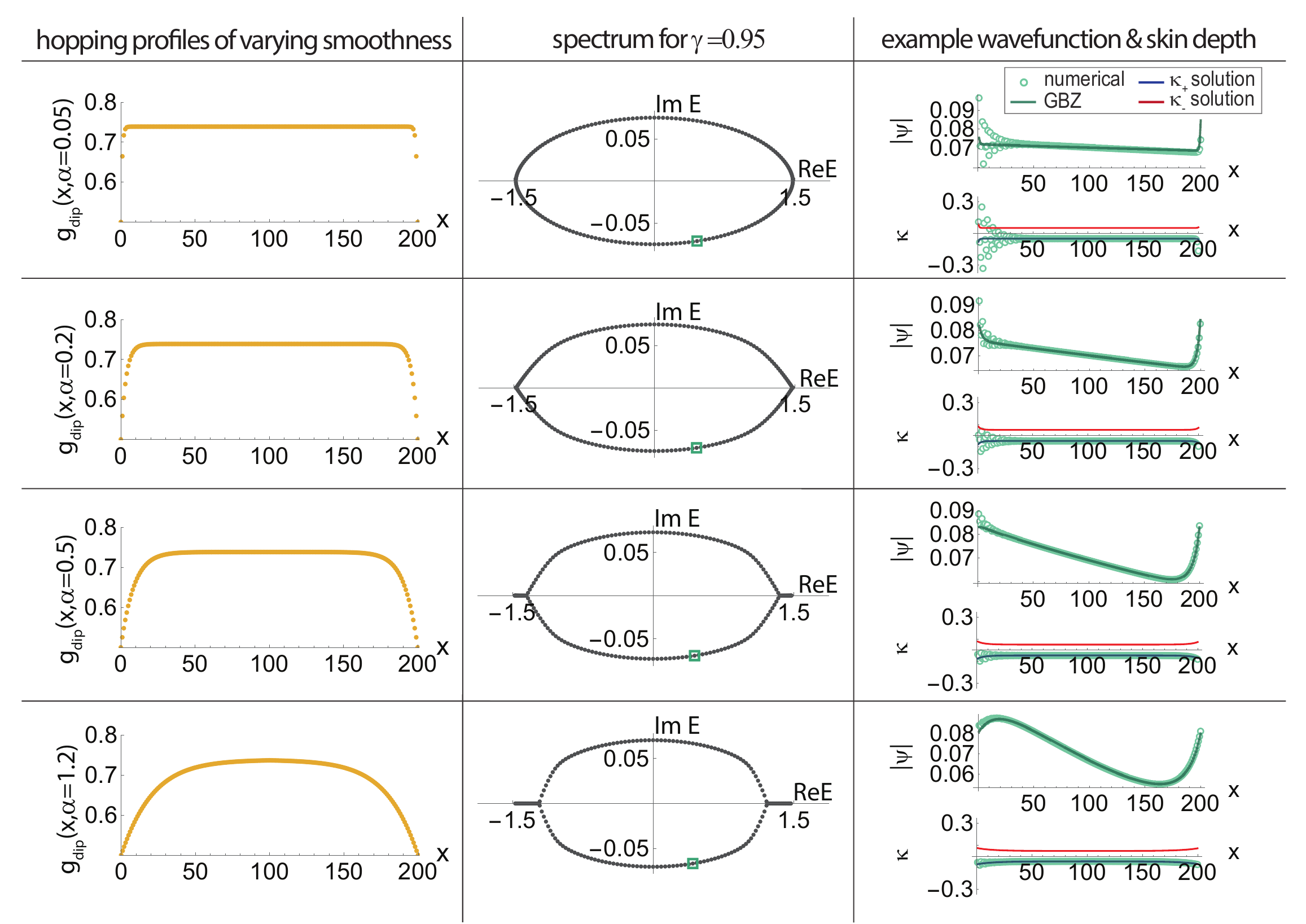}
	\caption{\textbf{Phase-space GBZ: effect of discontinuities in the hopping profile $g(x)$. }
 Numerical reconstruction of the phase-space GBZ becomes more noisy with increasing local non-smoothness in $g(x)$. The cases plotted are based on $g_{\text{dip}}(x,\alpha)=10^{-3}+(\tanh(\frac{2 \pi x}{\alpha L} + 1)-\tanh(1))$ for $-L/2<x\leq L/2$ [Eq.~\ref{NS_exp}], with varying smoothness parameter $\alpha=0.05, 0.2,0.5,1.2$ and fixed $L=200$, $\gamma=0.95$. A small $\alpha$, corresponding to a sharp discontinuity in $g(x)$, gives rise to significant Gibbs-like wavefunction fluctuations and hence fluctuating numerically reconstructed GBZ (green). Increasing $\alpha$ gradually suppresses these fluctuations until a smooth hopping profile and illustrative wavefunction are achieved. The phase-space GBZ, which is always a smooth curve, does not capture local fluctuations but consistently represents the best averaged trend, providing an optimal smoothed-fit for the wavefunctions.
 }
	\label{fig7B} 
\end{figure*}

\subsection{Limitations on the phase-space GBZ approach}

Although the phase-space GBZ has been shown to be effective in describing the spectra of inhomogeneous single- and double-component systems, local deviations, fluctuations, or discrepancies may still be observed for certain states, as exemplified in Fig.~\ref{0908fig3}c1 and c2, particularly near non-smooth regions of $g(x)$ or at the transition of $\kappa(x)$ into the pure skin region. In this section, additional examples of phase-space GBZs in single-component systems are presented, and the limitations of the phase-space GBZ method in various finite-size physical systems are examined.

We first showcase additional examples of two contrasting inhomogeneous $g(x)$ hopping profiles and their phase-space GBZ approximations in Fig.~\ref{fig7A}, such as to validate the accuracy of our approach. The two illustrative systems are:
{\small
\begin{align}
        &g_{\text{soft}}(x)= \begin{cases}
            5\tanh(\pi \frac{x}{L}+1)+10&\text{ for }x< \frac{2L}{5}\\
            5\tanh(\pi \frac{2}{5}+1)+10&\text{ for }\frac{2L}{5}\leq x\leq \frac{3L}{5}\\
            5\tanh(\pi-\frac{\pi x}{L}+1)+10&\text{ for }x> \frac{3L}{5}
        \end{cases},\label{decay1_exp}\\
        &g_{\text{bump}}(x)=1 + \sin(\pi x/L)^2 \cos(2 \pi x/L)^2,\label{osci_exp}
    \end{align}}
both with $\gamma=0.7$.

Their full spectra and illustrative states are well approximated by the phase-space GBZ compared to numerical results (Figs.~\ref{fig7A}b,~\ref{fig7A}c,~\ref{fig7A}e,~\ref{fig7A}f), with minor but observable localized fluctuations: slight deviations appear near the non-smoothness at $x=0$ and $x=L$ in the hopping profile $g(x)=g_{\text{soft}}(x)$ (Fig.~\ref{fig7A}a) as shown in Fig.~\ref{fig7A}c2; fluctuations also occur in the pure skin regions of states with coexisting pure and inhomogeneous skins (Figs.~\ref{fig7A}c2,~\ref{fig7A}c4,~\ref{fig7A}f4), where these GBZ $\kappa(x)$ fluctuations lead to localized deviations in the corresponding state profiles. We will elaborate on the reasons and the degree of effect of these two sources of inaccuracy later in this section.

\subsubsection{Non-smoothness in the hopping function}

In introducing the phase-space GBZ method, we assume sufficiently smooth hopping functions in Eq.~\ref{SmthApprox} such that $g(x\pm 1)$ is approximated as $g(x)$, leading to the local continuity of the continuous phase-space GBZ.

While Eq.~\ref{SmthApprox} remains valid in the thermodynamic limit $L\to\infty$ for any continuous hopping function $g(x)$ with $g(x+L)=g(x)$, the approximation precision diminishes for finite $L$ due to the local non-smoothness of $g(x)$. To illustrate the extent and characteristics of this diminishing accuracy, consider a hopping function with strongly decaying boundary hoppings:
{\small
\begin{align}
    g_{\text{dip}}(x,\alpha)&=\frac{1}{2}-\tanh(1)\notag\\
    &+\begin{cases}
        \tanh(\frac{2\pi x}{\alpha L}+1) &\text{ for }0<x\leq \frac{L}{2}\\
        \tanh(\frac{2\pi(L-x)}{\alpha L}+1) &\text{ for }\frac{L}{2}<x\leq L
    \end{cases},\label{NS_exp}
\end{align}}

as shown in Fig.~\ref{fig7B}. With $L=200$, the smoothness $\alpha$ is varied from $0.05$ to $1.2$, and the phase-space GBZ approximation is compared to numerical results.

With fixed $g_{\text{min}}$ and $g_{\text{max}}$, a reduction in $\alpha$ makes the violation of Eq.~\ref{SmthApprox} more pronounced, amplifying fluctuations in the wavefunction near the non-smooth regions of $g(x)$. Although the phase-space GBZ assumes perfectly smooth hopping functions and cannot fully capture these fluctuations, it effectively approximates the overall trend of the wavefunction, as demonstrated in Fig.~\ref{fig7B}.

When local fluctuations in the wavefunctions are of interest, it becomes necessary to enhance the smoothness of $g(x)$, most easily done by just increasing the number of sites within a fixed real-space length. In the special case of discontinuous $g(x)$ with sharp jumps over a small range, our phase-space GBZ can be modified to accommodate sharp boundaries, as discussed later in Sect.~\ref{sec_discont_gx}.

\subsubsection{Pure skin fluctuation in finite-size systems}\label{pure_fluc}

In the phase-space GBZ approach, inhomogeneous systems are approximated by assuming the thermodynamic limit $L\to \infty$, where the bulk relation becomes fully decoupled between neighbouring sites:
\begin{align}
    &\beta_{n}(x)+\frac{1}{\beta_{n}(x)}=\frac{E_{n}}{g(x)}.
\end{align}
Within this framework, discontinuous phase-space GBZ states, which have real eigenenergies and consist of both the pure skin and inhomogeneous skin regions, have $\kappa_{n,\pm}(x)$ that goes from being zero to non-zero at the boundary between pure skin and inhomogeneous skin regions.

In practice, with finite $L$, neighbouring $\beta_n(x)$ does differ slightly, necessitating the use of the original bulk equation [Eq.~\ref{BulkApprox}] without the approximation $\beta_n(x+1)\approx \beta_n(x)$.

This then brings in problems at the pure-inhomogeneous skin boundaries, where neighbouring phase-space GBZ factors $\beta_{n,\pm}(x)$ and $\beta_{n,\pm}(x+1)$ are located within different skin regions, and only one of them is complex. This is mathematically prohibited in Eq.~\ref{BulkApprox} for real $E_n$ and $g(x)$. A necessarily non-zero imaginary component of $\beta_{n}(x)$ of pure skin, or equivalently $\kappa_{n}(x)\neq 0$ in the pure skin regions, is required to balance the imaginary equation of Eq.~\ref{BulkApprox} in finite-size systems.

Nevertheless, the phase-space GBZ solutions $\kappa_{n,\pm}(x)=0$ still provide an effective best-fit approximation for the fluctuating $\kappa_{n}$. Moreover, despite the fluctuations in $\kappa_{n}(x)$, their impact on the wavefunctions remains minimal, as demonstrated in examples of Figs.~\ref{fig7A}b,~\ref{fig7A}c,~\ref{fig7A}e,~\ref{fig7A}f, where the effect of GBZ fluctuations is restricted locally by the effectively zero skin in the pure skin regions and is compressed by the inhomogeneous skin, which amplifies the wavefunctions at one of the sides.

\begin{figure*}[!htbp]
    \includegraphics[width=0.8\linewidth]{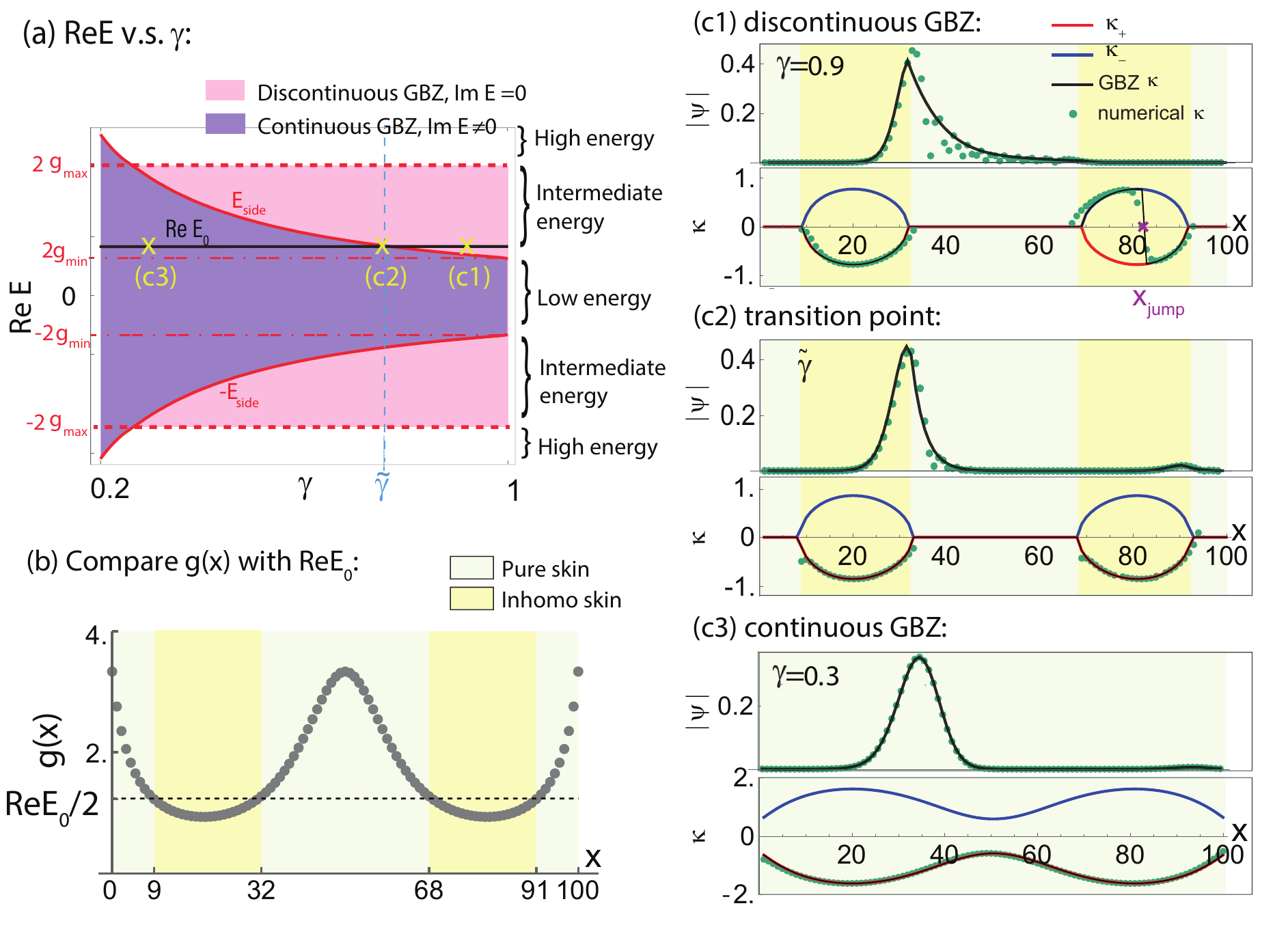}
	\caption{\textbf{$\gamma$-driven transition between continuous and discontinuous phase-space GBZs. }
(a) The phase diagram of a 1-component lattice with $g(x)=1/(\sin(2 \pi x/L) \cos(\pi x/L) + 0.3)$ and $L=100$, in the Re$E$-$\gamma$ parameter space. The curves $\pm E_{\text{side}}$, as calculated from Eq.~\ref{sideReE2}, separate the continuous and discontinuous phase-space GBZ phases.
For a fixed chosen value of Re$E_0$ (black), one can always find the threshold value $\gamma=\Tilde{\gamma}$ [Eq.~\ref{sideReE}] where the transition occurs. Here, $E_0=E_{\text{side}}=2.46$ when $\gamma$ is tuned to the value $\Tilde{\gamma}=0.76$.
(b) For a eigenstate corresponding to $E_0$, its pure/inhomogeneous skin regions correspond to positions where $2g(x)$ is above/below the value of Re$E_0$.
(c) Illustrative states at various $\gamma$, correspond to (c1) the discontinuous GBZ phase, (c2) at the transition point and (c3) in the continuous GBZ phase. Clear-cut regions of pure and inhomogeneous skins exist in the discontinuous GBZ and the transition point states. Within the inhomogeneous skin region, the local continuity of the GBZ ensures numerical $|\kappa(x)|$ consistently adheres to one of $|\kappa_{\pm}(x)|$, even near the sign-inversion site. This enables the calculation of the single $x_{\text{jump}}$ according to Eq.~\ref{GBZBCinv}. 
 }
	\label{fig3old} 
\end{figure*}

\subsection{Phase-space GBZ transitions for a fixed state}
\label{sec:single_state}
The energy eigenvalue $E_{\text{side}}$ [Eq.~\ref{sideReE2}] not only bounds the spectral loop on both sides but also serves a crucial role in confining the real ``tails'' within the intermediate energy regime, which is bounded by $\pm 2g_{\text{max}}$. In essence, $E_{\text{side}}$ acts as a transition point between continuous and discontinuous phase-space GBZ, and is calculated from the boundary requirement Eq.~\ref{sideReE2} (Eq.~\ref{2BspecGBZ}) in single-component (double-component) chains, given known non-Hermiticity $\gamma$ ($\gamma_1 \gamma_2$) and hopping function $g(x)$ ($g_{1,2}(x)$).

Here, we elaborate on how one can alternatively investigate the phase-space GBZ by examining how a specific $E_0$ state undergoes transitions as a certain parameter is varied. For instance, in the topological investigation of the zero-energy state $E_{0}=0$, the non-Hermiticity $\gamma$ is tuned such that the state undergoes a transition between topologically trivial and non-trivial phases. We hereby propose the concept of threshold value $\Tilde{\gamma}$ of $\gamma$, defined for a fixed specified $E_0$. In a single-component system, it is given by
\begin{align}
   \Tilde{\gamma}(E_0)=&\exp\left(-\frac{1}{L}\sum_{x=1}^{L} \cosh^{-1}\left(\frac{\text{Re }E_{0}}{2g(x)}\right)\right).\label{sideReE}
\end{align}
It gives the value of $\gamma$ across which a phase transition between continuous and discontinuous phase-space GBZ occurs, as demonstrated in Fig.~\ref{fig3old}. The state $E_0$ is real and resides within the discontinuous GBZ phase if $|\log(\gamma)|\leq|\log(\Tilde{\gamma})|$, but has to be complex otherwise.

Fig.~\ref{fig3old}a shows the phase diagram of state $E_0$ in the Re$E$-$\gamma$ space, where $\Tilde{\gamma}(E_0)$ is determined as the point where Re$E_0$ becomes the $E_\text{side}$. For a fixed value of Re$E_0$, $\Tilde{\gamma}(E_0)$ separates the (c3) continuous (complex $E_0$) and (c1,c2) discontinuous (real $E_0$) phases.  Crucially, the real state $E_0$ within the discontinuous GBZ phase, and consequently its phase-space GBZ solutions $\kappa_{n,\pm}$, remain robust against slight variation in $\gamma$, as verified by comparing (c1) and (c2). Adjusting $\gamma$ solely shifts the discontinuity position $x_{\text{jump}}$ along the real-space position.

A similar threshold non-Hermiticity $\Tilde{\gamma}$ is defined for 2-component states with energy $E_0$ within the lowest energy regime or the intermediate energy regime:
{\small
\begin{align}
    &\Tilde{\gamma}(E_0)=\exp\left(-\frac{1}{L}\sum_{1<|\omega_{ j }(\text{Re} (E_0),x)|}\cosh^{-1}|\omega_{ j }(\text{Re}(E_0),x)|\right),
\end{align}}
where $\Tilde{\gamma}_{\text{topo}}=\Tilde{\gamma}(0)$ sets the threshold non-Hermiticity for the topological zero mode.

\begin{figure*}[!htpb]
    \centering
    \includegraphics[width=0.9\linewidth]{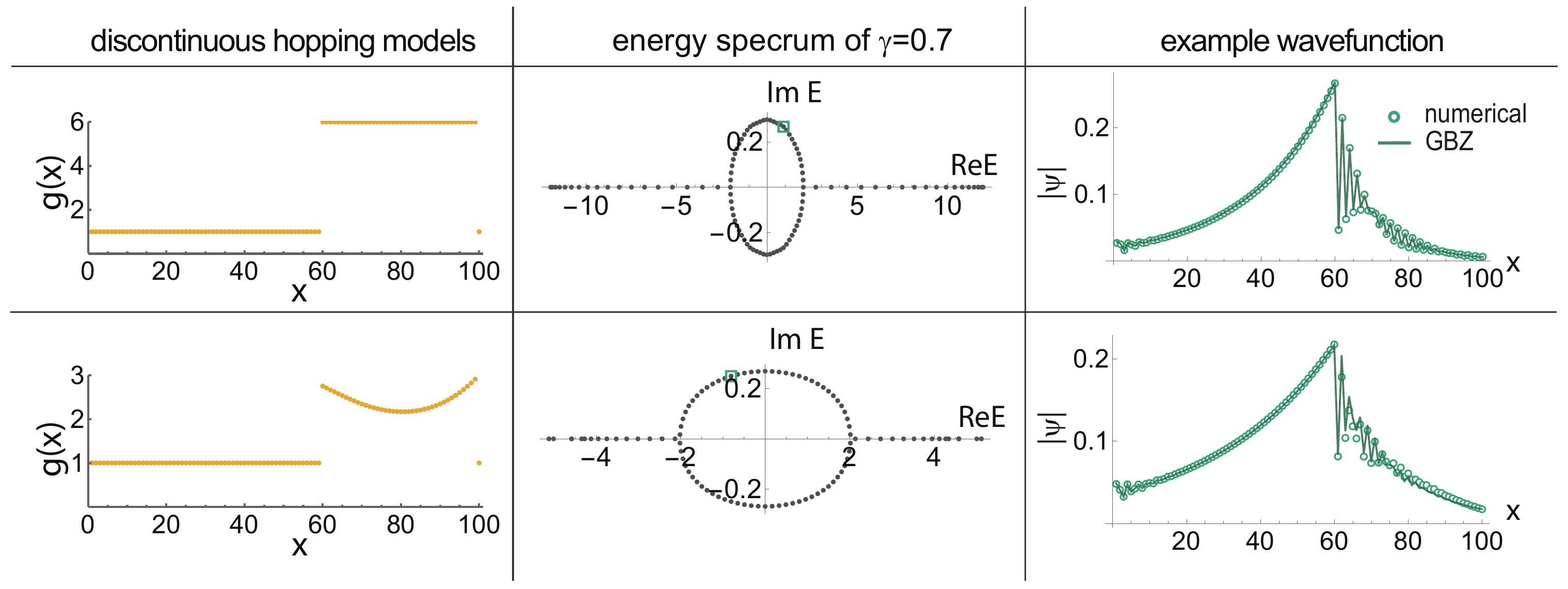}
    \caption{\textbf{Incorporating isolated discontinuities in the hopping profiles into the phase-space GBZ. } Although the accuracy of the phase-space GBZ approach may be reduced when non-smoothness of $g(x)$ is introduced, isolated non-smoothness or discontinuities can be effectively addressed by matching superposed solutions at the discontinuity sites, as given by Eq.~\ref{NS3}. (a) Two illustrative hopping profiles with discontinuities at $x=0$ and $\frac{3L}{5}$, both with $g(x)=1$ for $0\leq x < \frac{3L}{5}$. For the remaining lattice, the upper example has $g(x)=6$ and the lower has $g(x)=1/(\sin(2 \pi x/L) \cos(\pi x/L) + 2)$. (b) Their spectra at $\gamma=0.7$, both which exhibit characteristic real tails. (c) By applying Eq.~\ref{NS3} at $x=0$ and $\frac{3L}{5}$ for two illustrative states (indicated by green hollow squares in (b)), the resultant wavefunction prediction (solid line) is seen to agree well with the numerical wavefunctions (circles) for both $g(x)$ profiles.} 
    \label{fig10} 
\end{figure*}

\subsection{Handling isolated hopping discontinuities}\label{sec_discont_gx}
Here we extend our phase-space GBZ formalism to handle systems with local isolated non-smoothness or discontinuity in the hoppings, such that they are treated as additional boundaries. 
Consider a non-Hermitian inhomogeneous lattice with a discontinuous hopping function
\begin{align*}
    &g(x)=
    \begin{cases}
      f_{1}(x) & \text{ for } x_{0}\leq x< x_{1}\\
      f_{2}(x) & \text{ for } x_{1} \leq x< L+x_{0}
    \end{cases},
\end{align*}
where $f_{1,2}(x)$ are smooth functions with $f_{1}(x_1-1)\neq f_2(x_1)$. Its eigenstate solution $\psi(x)$ is solved by separately considering $\psi^{(1,2)}(x)$:
\begin{align}
&\psi(x)=
\begin{cases}
     \psi^{(1)}(x) & \text{ for } x_{0}\leq x\leq x_{1}\\
     \psi^{(2)}(x) & \text{ for } x_{1} \leq x\leq L+x_{0}
\end{cases},
\end{align}
with two boundary conditions at each discontinuous point:
 \begin{align}
    &\psi^{(1)}(x_{0})=\psi^{(2)}(x_{0}),\\
    &f_{1}(x_{0})\psi^{(1)}(x_{0}+1)=f_{2}(x_{0})\psi^{(2)}(x_{0}+1).\label{NSdeal}
\end{align}
$\psi^{(1,2)}(x)$ could be written as a superposition of phase-space GBZ state solutions of \emph{both} smooth-hopping sub-lattices simultaneously:
\begin{align}
    &\psi^{(1,2)}(x)=c_{+}^{(1,2)}\phi_{+}^{(1,2)}(x)+c_{-}^{(1,2)}\phi_{-}^{(1,2)}(x),\label{NS1}
\end{align}
with
\begin{align}
     &\phi_{\pm}^{(1,2)}(x)= \frac{1}{\sqrt{f_{1,2}(x)}}\gamma^x\prod_{x'=1}^{x}\beta_{n,\pm}^{(1,2)}(x).
\end{align}
The superposed state solutions fulfill the bulk relation:
{\small
\begin{align}
f_{1,2}(x)&/\gamma\psi^{(1,2)}(x+1)+f_{1,2}(x-1) \gamma\psi^{(1,2)}(x-1)=E_{n},\label{jumpSubBulk}
\end{align}}
for all $x$ defined on $f_{1,2}(x)$. In contrast to smooth-hopping states that have only one of $\beta_{n,\pm}(x)$ contributing to the wavefunction, each non-smooth jump in $g(x)$ requires simultaneous contributions from both solutions, where one governs the overall trend and the other accounts for the fluctuation near the jump. Generally, two sub-lattices with two non-smooth jumps involve four non-trivial coefficients $c_{\pm}^{(1,2)}$ to be solved: 
\begin{align}
    \text{Det}\begin{bmatrix}
c_{+}^{(1)} & c_{-}^{(1)} & -c_{+}^{(2)} & -c_{-}^{(2)} \\
F_{1,+,+}& F_{1,-,-}& -F_{2,+,+}&-F_{2,-,-}\\
U_{1,+,+}& U_{1,-,-}& -U_{2,+,+}& -U_{2,-,-}\\
W_{1,+,+}&W_{1,-,-} &-W_{2,+,+}& -W_{2,-,-}
    \end{bmatrix}=0,\label{NS2}
\end{align}
where 
\begin{align}
     F_{a,\mu,\nu}&=f_{a}(x_{0})c_{\mu}^{(a)}\gamma\beta_{n,\nu}^{(a)}(x_{0}+1),\\
U_{a,\mu,\nu}&=c_{\mu}^{(a)}\prod_{x=x_{0}+1}^{x_{1}}\gamma\beta_{n,\nu}^{(a)}(x),\\
W_{a,\mu,\nu}&=f_{a}(x_{1})c_{\mu}^{(a)}\prod_{x=x_{0}+1}^{x_{1}+1}\gamma\beta_{n,\nu}^{(a)}(x).
\end{align}
In practice, with large $L$, strong NHSE accumulation occurs in one of the sub-lattices due to the existence of inhomogeneous skin regions:
\begin{align}
    &\prod_{x=x_{0}+1}^{x_{1}}\gamma\beta_{n,-}^{(j)}(x)\ll\prod_{x=x_{0}+1}^{x_{1}}\gamma\beta_{n,+}^{(j)}(x),\label{NS4}
\end{align}

and Eq.~\ref{NS2} can be simplified by the approximation $c_{-}^{(j)}\to 0$, which is justified by Eq.~\ref{NS4} . For $j=1$, Eq.~\ref{NS2} reduces to
\begin{align}
    \text{Det}\begin{bmatrix}
        U_{2,+,+}+U_{1,+,+}  & U_{2,-,-}\\
            W_{2,+,+}+W_{1,+,+} &W_{2,-,-}
    \end{bmatrix}=0,\label{NS3}
\end{align}

whose solution gives all $c_{-}^{(j)}=0$ and completes the phase-space GBZ construction for systems with isolated non-smoothness. Fig.~\ref{fig10} demonstrates examples of $g(x)$ with a pair of discontinuities and how their states are well approximated by the GBZ approach with an additional discontinuity [Eq.~\ref{NS3}].

\bibliography{references_inhomo}
\bigskip

\end{document}